\documentclass[12pt]{article}

\usepackage{amssymb}
\usepackage{amsmath}
\usepackage{amscd}
\usepackage{latexsym}
\usepackage{graphicx}

\usepackage{cite}
\usepackage{enumitem}

\newcommand{\be}{\begin{equation}}
\newcommand{\ee}{\end{equation}}

\newcommand{\prt}{\partial}

\newcommand{\bt}{\beta}
\newcommand{\vp}{\varphi}
\newcommand{\ep}{\varepsilon}
\newcommand{\al}{\alpha}

\newcommand{\gm}{\gamma}

\newcommand{\Gm}{\Gamma}
\newcommand{\ra}{\rightarrow}
\newcommand{\cA}{{\cal A}}
\newcommand{\lbd}{\lambda}

\begin{document}

\begin{center}

{\Large{\bf Self-similar extrapolation in quantum field theory} \\ [5mm]

V.I. Yukalov$^{1,2,*}$ and E.P. Yukalova$^3$} \\ [3mm]

$^1${\it Bogolubov Laboratory of Theoretical Physics, \\
Joint Institute for Nuclear Research, Dubna 141980, Russia \\ [2mm]

$^2$Instituto de Fisica de S\~ao Carlos, Universidade de S\~ao Paulo, \\
CP 369,  S\~ao Carlos 13560-970, S\~ao Paulo, Brazil   \\ [2mm]

$^3$Laboratory of Information Technologies, \\
Joint Institute for Nuclear Research, Dubna 141980, Russia } \\ [3mm]

$^*${\it Corresponding author e-mail}: yukalov@theor.jinr.ru \\ [3mm]

\end{center}

\vskip 2cm

\begin{abstract} 
Calculations in field theory are usually accomplished by employing some variants of 
perturbation theory, for instance using loop expansions. These calculations result 
in asymptotic series in powers of small coupling parameters, which as a rule are 
divergent for finite values of the parameters. In this paper, a method is described 
allowing for the extrapolation of such asymptotic series to finite values of the 
coupling parameters, and even to their infinite limits. The method is based on 
self-similar approximation theory. This theory approximates well a large class of 
functions, rational, irrational, and transcendental. A method is presented, resulting 
in self-similar factor approximants allowing for the extrapolation of functions to 
arbitrary values of coupling parameters from only the knowledge of expansions in 
powers of small coupling parameters. The efficiency of the method is illustrated by 
several problems of quantum field theory.
 
\end{abstract}

\vskip 1cm

{\parindent=0pt

}

\newpage

\section{Introduction}

The solution of almost all nontrivial problems resorts to the use of some kind of
perturbation theory yielding asymptotic series in powers of small parameters. 
However the physical values of the parameters of interest are usually not small and
often are even quite large. Thus we come to the necessity of being able to extrapolate
the asymptotic series, that are usually divergent, to the finite values of the 
parameters of interest. Moreover, sometimes the main interest is in the behaviour of
the studied characteristics at asymptotically large parameters tending to infinity. 
Pad\'{e} approximants can sometimes extrapolate small-variable series to the 
finite-variable region. However, as is well known, they cannot describe the 
large-variable behaviour at the variable tending to infinity, if only a small-variable
expansion is available \cite{Baker_1}. Let us emphasize that here we keep in mind 
the case where no large-variable behaviour is known, because of which it is impossible
to turn to two-point Pad\'{e} approximants requiring the knowledge of the large-variable
behaviour \cite{Baker_1,Baker_2}. Similarly, it is not possible to use other 
interpolation methods needing the information on the large-variable asymptotic 
behaviour for the quantity of interest. Our aim here is to consider not interpolation 
but {\it extrapolation}, when only the small-variable expansions are available. 

Pad\'{e} approximants, as is known, provide the best approximation for rational 
functions, but the reason why they cannot predict the large-variable behavior for 
irrational functions is rather straightforward. Really a $P_{M/N}(x)$ Pad\'{e} 
approximant in the limit of $x \ra \infty$ behaves as $x^{M-N}$, where $M$ and $N$ 
are integers. Moreover, the difference $M-N$ depends on the used Pad\'{e} approximant, 
but not uniquely defines the limiting exponent.  

The other method of extrapolation, Borel summation, requires the knowledge of the
large-order behaviour of series coefficients, so that the error of the truncated series 
be bounded by $C^n n! |z|^n$, where $C$ is a constant \cite{Hardy_3,Kleinert_4,Weinberg_5}. 
However, this large-order behavior of the expansion coefficients not always is known. 
There exist several variants of the approach involving Borel summation, including the 
combination of the Borel transform, conformal mapping, and Pad\'{e} approximants
\cite{Hardy_3,Kleinert_4,Weinberg_5,Costin_50,Costin_51,Costin_52}. 

Among other methods allowing for the large-variable extrapolation, it is possible to 
mention the approach based on the introduction of control functions defined by 
fixed-point conditions optimizing the series convergence \cite{Yukalov_53,Yukalov_54}. 
Several variants of this approach have been considered, e.g., \cite{Yukalov_53,Yukalov_54,
Yukalov_55,Yukalov_56,Caswell_57,Halliday_58,Killingbeck_59,Stevenson_60,Feranchuk_61,
Yukalov_62,Dineykhan_63,Sissakian_64,Kleinert_65,Feranchuk_66}. The introduction of 
control functions requires to rearrange the considered series by either a change of
the variable containing trial parameters \cite{Kleinert_4} or by incorporating trial 
parameters into an initial approximation \cite{Yukalov_53,Yukalov_54}.    

The series convergence depends on the choice of an initial approximation. In many cases, 
one chooses as an initial approximation a Gaussian form corresponding to free particles. 
More complicated forms for the initial approximation can also be chosen. For example, one
can start perturbation theory with a non-Gaussian approximation 
\cite{Shaverdyan_67,Ushveridze_68,Turbiner_69,Sazonov_70} or one can use for the initial 
approximation nontrivial Hamiltonians, as in the method of Hamiltonian envelopes 
\cite{Yukalov_71}.   

The methods mentioned above are numerical, require rearrangements of perturbation 
series and rather involved calculations. It would be good to have a simple analytical 
method that could extrapolate the standard Taylor series derived by the usual perturbation 
theory in powers of a parameter, say the coupling parameter.   
 
Here we describe such a simple general method allowing for the extrapolation of asymptotic 
series in powers of a small variable to arbitrary values of this variable, including 
infinity. As illustration, we accomplish the extrapolation of the series for several
functions met in quantum field theory, whose behaviour at large coupling parameters is 
of interest by its own. The advantages of the suggested approach, as compared to other
methods, are as follows.

(i) First of all, the suggested method is analytical allowing for the derivation of 
explicit forms of the sought functions. This makes it straightforward to analyse the 
results with respect to different parameters entering the problem, which is not always 
easy in numerical methods.

(ii) Moreover, numerical methods in some cases are not applicable, while the presented 
method of extrapolation of asymptotic series can always be applied, provided at least 
several terms of perturbation theory are available.   

(iii) Even if numerical simulations could be invoked, they usually require powerful 
computational facilities and essential calculational time. On the contrary, the suggested 
method is very simple and straightforward. 
      
(iv) Numerical methods have their own problems and limitations. Therefore the employment 
of simpler analytical methods can serve as a guide for numerical calculations.     

(v) Finally, the physics of the considered problem becomes much more transparent when
possessing an explicit, although approximate, formula allowing for studying its behavior
in limiting cases.

It is important to stress that the main aim of the article is to develop a method of 
extrapolation that would be simple, analytical, and applicable even for those cases 
where just a few terms of perturbative series are available. This is why we concentrate 
our attention on these points.   

When numerous terms of a series are available, there are several methods allowing for 
accurate extrapolation. However this is not the point of our interest. If we were 
interested in getting high accuracy of extrapolation for series with numerous terms, 
we should resort to some modifications of our method, using additional tricks, such as 
the introduction of control functions into self-similar approximants \cite{Kleinert_65}, 
the combination of self-similar approximations with Pade approximants or with Borel 
transforms, etc. Thus it has been demonstrated \cite{Gluzman_72,Gluzman_73} that the 
combination of self-similar and Pad\'{e} approximants converges much faster and provides 
essentially higher accuracy then the best Pad\'{e} approximants of the same order. 
However all that is a quite different problem requiring separate investigations and 
publications, some of which we cite. We stress it again that the main aim of the present
paper is to suggest a simple analytical method providing reliable approximations when 
other methods are not applicable.

\section{Self-similar approximation theory}

The suggested method is based on self-similar approximation theory advanced in 
Refs. \cite{Yukalov_6,Yukalov_7,Yukalov_8,Yukalov_9,Yukalov_10}. In Ref. \cite{Yukalov_15},
this theory is used for developing a convenient approach to the problems of 
{\it interpolation} in high-energy physics, when weak-coupling as well as strong-coupling 
expansions are known. Here we extend the applicability of the approach for the essentially 
more complicated problem of {\it extrapolation} in quantum field theory, when only the 
weak-coupling asymptotic series are available, but the behavior at the strong-coupling
limit is not known, and even more, finding this behavior is the point of main interest.       

First we briefly recall the main ideas of self-similar approximation theory in order that 
the reader could understand its justifications and would get the feeling why it can 
successfully work. This approach is based on mathematical techniques of renormalization 
group theory, dynamical theory, and optimal control theory 
\cite{Foulds_11,Hocking_12,Yukalov_13}. Note that these theories are closely interrelated 
since the renormalization group theory, actually, is a particular case of dynamical theory.  

The pivotal idea is to reformulate perturbation theory to the language of dynamical theory 
or renormalization-group theory. For this purpose, we treat the approximation-order index 
as discrete time and the passage from one approximation to another as the motion in the 
space of approximations. Suppose, we can find the sought function only as a sequence of 
approximations at a small variable, $f(x) \simeq f_k(x)$ for $x \ra 0$, where 
$k = 0, 1, 2, \ldots$ is the approximation order. For concreteness, we consider here 
real-valued functions of real variables. The extension to complex-valued functions can be 
straightforwardly done by considering several functions corresponding to real and imaginary 
parts of the sought function.  

The sequence of the bare approximants $f_k(x)$ is usually divergent. Therefore the 
first thing that is necessary to do is to reorganize this sequence by introducing 
control functions $u_k = u_k(x)$ governing the sequence convergence. Control functions 
can be incorporated in several ways, through initial conditions, calculational algorithm, 
or by a sequence transformation. Thus, instead of the bare approximants $f_k(x)$, 
we pass to a transformed sequence of the approximants 
$$
F_k(x,u_k) = \hat T[u] f_k(x) \; .
$$ 
For short, we write here one control function $u_k$, although there can be several 
of them, so that $u_k$ can be understood as a set of the necessary control functions. 
We assume that the used transformation is invertible, in the sense that 
$$
f_k(x) = \hat T^{-1}[u] F_k(x,u_k) \; . 
$$

The sequence $\{F_k(x,u_k)\}$ is convergent if and only if it satisfies the Cauchy 
criterion, when for each $\varepsilon > 0$ there exists a number $k_c$ such that 
$$
|F_{k+p}(x,u_{k+p}) - F_k(x,u_k)| < \ep
$$ 
for all $k > k_c$ and $p > 0$. In the language of optimal control theory, this implies 
that control functions can be defined as the minimizers of the convergence cost functional 
\cite{Yukalov_13}
\be
\label{1}
 C[u] = \sum_k | F_{k+1}(x,u_{k+1}) - F_k(x,u_k) | \;  .
\ee

In order to formulate the passage between different $F_k$ as the evolution of a dynamical 
system, it is necessary to define an endomorphism in the space of approximants 
$$
\cA = \{ F_k(x): \; k = 0,1,2,\ldots;\; x\in \mathbb{R} \} \; .
$$
For this purpose, we introduce the expansion function $x = x_k(f)$ by the reonomic 
constraint 
$$
F_0(x,u_k(x)) = f \; .
$$
The endomorphism in the approximation space is defined as
$$
y_k(f) \equiv F_k(x_k(f), u_k(x_k(f))) \; ,
$$ 
with the inverse relation 
$$
F_k(x,u_k(x) ) = y_k(F_0(x,u_k(x))) \; .
$$

By this construction, the approximation sequence $\{F_k\}$ is bijective with the
sequence of the endomorphisms $\{y_k\}$. Therefore, if the sequence $\{F_k\}$ 
converges to a limit $F^*$, then the sequence of the endomorphisms $\{y_k\}$ 
converges to a limit $y^*$. The limit $y^*$ plays the role of a fixed point for the 
endomorphism sequence $\{y_k\}$, where $y_k(y^*(f) ) = y^*(f)$. In the vicinity of a 
fixed point, the endomorphism enjoys the property of self-similarity
\be
\label{2}
 y_{k+p}(f) = y_k(y_p(f) ) \; ,
\ee
with the initial condition $y_0(f) = f$. This is, actually, just the semi-group property 
$y_{k+p} = y_k \cdot y_p$, with the unity element $y_0 = 1$. The sequence of endomorphisms, 
with the above semi-group property, is called cascade (or semi-cascade),
$$
\{ y_k(f): \; \mathbb{Z}_+ \times \mathbb{R} ~ \ra ~ \mathbb{R} \} \; , 
$$
where the role of time is played by the approximation order $k$. 

A cascade , which is a dynamical system in discrete time, can be embedded \cite{Clifford_14} 
into a flow that is a dynamical system in continuous time, 
$$
\{ y_k(f): \; \mathbb{Z}_+ \times \mathbb{R} \ra \mathbb{R} \} \subset
\{ y(t,f): ~ \mathbb{R}_+ \times \mathbb{R} ~ \ra ~ \mathbb{R} \}    \, .
$$
The embedding implies that the flow enjoys the same group property 
$$
y(t+t',f) = y(t,y(t',f))
$$ 
and the flow trajectory passes through all points of the cascade trajectory, 
$$
y(t,f) = y_k(f) \qquad (t = k) \; , 
$$
with the same initial condition $y(0,f) = f$.

The above group property can be rewritten as the Lie differential equation 
\be
\label{3}
 \frac{\prt}{\prt t} \; y(t,f) = v(y(t,f) ) \; ,
\ee
with the velocity 
$$
v(y) \equiv \lim_{\tau \ra 0} \; \frac{\prt}{\prt\tau} \; y(\tau,y) \; .
$$
Integrating the differential evolution equation (\ref{3}) yields the evolution integral
\be
\label{4}
 \int_{y_k}^{y_k^*} \frac{dy}{v(y)} = t_k \; ,
\ee
in which the integration is from a point $y_k(f)$ to an approximate fixed point 
$y_k^*(f)$, with $t_k$ being the effective time needed for reaching the latter point. 
Here $y_k^*(f)$ is an approximate fixed point, since in practice we always have to 
limit the consideration by a finite number of steps. Taking in the evolution integral 
the cascade velocity represented in the form of the Euler discretization 
$$
v_k(f) = y_{k+1}(f) - y_k(f) \; , 
$$
we come to the integral
\be
\label{5}
\int_{F_k}^{F_k^*} \frac{df}{v_k(f) } = t_k \; ,
\ee
in which $$
F_k^*(x) = y_k^*(F_0(x,u_k(x) )
$$ 
is the effective limit of the sequence $\{F_k\}$ corresponding to the approximate fixed 
point $y^*_k$. Applying the inverse transformation, we obtain the self-similar approximant
$$
f_k^*(x) = \hat T^{-1}[u] F_k^*(x) \; . 
$$ 
  
These are the principal steps in deriving self-similar approximants. The practical 
realization depends on the form of the bare approximants $f_k(x)$, the concrete form 
of the transformation $\hat{T}$ introducing control functions, and on the method of 
defining the latter. When the asymptotic behaviour of the sought quantity is known for
small as well as for large coupling parameters, it is convenient to accomplish the 
interpolation with the use of self-similar root approximants, as is demonstrated in
Ref. \cite{Yukalov_15}. But for the problem of {\it extrapolation} we need to employ 
another type of approximants.    

Usually, the asymptotic behaviour at small coupling parameters $x \ra 0$, is of the form
\be
\label{6}
f_k(x) = f_0(x) \left( 1 + \sum_{n=1}^k a_n x^n \right ) \;   ,
\ee
where $f_0(x)$ is a given function. The  above sum is usually divergent for finite values 
of $x$, hence makes no sense for finite $x$. Moreover, often it is necessary to find the 
behaviour of the sought function $f(x)$ at asymptotically large $x \ra \infty$. 

By the fundamental theorem of algebra \cite{Lang_16}, a polynomial of any degree of 
one real variable over the field of real numbers can be split in a unique way into a 
product of irreducible first-degree polynomials over the field of complex numbers. This 
implies that the finite series (\ref{6}) can be represented as the product
\be
\label{7}
 f_k(x) = f_0(x) \prod_j \left( 1 +  b_j x \right ) \; ,
\ee
with $b_j$ expressed through $a_n$.  

Control functions can be explicitly incorporated by employing fractal transforms 
\cite{Yukalov_13,Barnsley_17}, which can be written in the form
\be
\label{8}
 \hat T[s,u] f_k(x) = x^{s_k} f_k(x) + u_k \; .
\ee
Then, following the scheme described above, we obtain the self-similar factor approximants
\be
\label{9}
f_k^*(x) = f_0(x) \prod_{j=1}^{N_k} \left( 1 +  A_j x \right )^{n_j} \;  ,
\ee
with $A_j$ and $n_j$ playing the role of control parameters 
\cite{Yukalov_18,Gluzman_19,Yukalov_20}.

The number of factors $N_k$ equals $k/2$ for even $k$ and $(k + 1)/ 2$ for odd $k$. 
A factor approximant (\ref{9}) represents the sought function, therefore their asymptotic 
expansions should coincide. Then the parameters $A_j$ and $n_j$ are to be chosen so that 
the asymptotic expansion of approximant (\ref{9}) of order $k$ be equal to the asymptotic 
form (\ref{6}), that is, $f_k^*(x) \simeq f_k(x)$ for $x \ra 0$. This condition yields 
the equations
\be
\label{10}
 \sum_{j=1}^{N_k} n_j A_j^n = D_n \qquad ( n = 1,2,\ldots, k) \;  ,
\ee
where
$$
D_n \equiv \frac{(-1)^{n-1}}{(n-1)!} \; \lim_{x\ra 0} \; 
\frac{d^n}{dx^n} \; \ln\left( 1 + \sum_{m=1}^n a_m x^m \right ) \; .
$$
 
When $k$ is even, hence $N_k = k/2$, we have $k$ equations for $k$ unknown parameters 
$A_j$ and $n_j$, uniquely defining these parameters \cite{Yukalov_20}. However if $k$ 
is odd, and $N_k = (k + 1)/2$, we have $k$ equations for $k + 1$ parameters. Then, 
to make the system of equations complete, it is necessary to add one more condition. 
For instance, resorting to scaling arguments \cite{Yukalov_20}, it is possible to set 
one of $A_j$ to one, say fixing $A_1 = 1$. This method gives for odd approximants 
the results close to the nearest even-order approximants. However below we prefer to
deal with uniquely defined even orders. Sometimes it may happen that the solutions for 
the parameters $A_j$ and $n_j$ are complex-valued. But this does not lead to any problem, 
since such complex solutions for the parameters appear in complex conjugate pairs, so 
that the whole expression remains real valued.     

When we are interested in predicting the large-variable behaviour of a function $f(x)$, 
we study the self-similar factor approximant (\ref{9}) at $x \ra \infty$. If the 
function $f_0$ behaves as $f_0(x) \simeq A x^\al$ for $x \ra \infty$, then the 
self-similar factor approximant (\ref{9}) for large $x$ is
\be
\label{11}
 f_x^*(x) \simeq B_k x^{\gm_k} \qquad ( x \ra \infty) \;  ,
\ee
with the amplitude and the exponent
\be
\label{12}
B_k = A \prod_{j=1}^{N_k} A_j^{n_j} \; , \qquad \gm_k = \al + \sum_{j=1}^{N_k} n_j \; .
\ee

In those cases, where the large-variable asymptotic behaviour of the sought function 
is known, say being $f(x) \simeq B x^\gm$ for $x \ra \infty$, it is straightforward 
to determine the accuracy of the prediction by calculating the percentage errors 
$\ep(B_k) \equiv ((B_k-B)/B) \times 100\%$ and 
$\ep(\gm_k) \equiv ((\gm_k-\gm)/\gm) \times 100\%$. When the exact large-variable 
asymptotic behaviour is not available, one usually presents the difference between 
the subsequent approximations for the quantity of interest. This difference 
characterizes the variation bar or dispersion of the obtained results, which is 
related to the stability of the calculational procedure \cite{Higham_78}.
If the subsequent results strongly differ from each other, this induces suspicion
of the procedure stability.  

Often, the most important hard case is the prediction of the exponent in the 
strong-coupling limit, since the value of the exponent essentially defines the physics 
of the problem. This hard case is the main study in the present article.   

It is important to stress that the derivation of self-similar factor approximants is
based on the Cauchy criterion of convergence, so that these approximants are expected 
to converge by construction. Moreover the numerical convergence of self-similar factor 
approximants has been confirmed by a number of problems enjoying many terms in their 
asymptotic expansions, when a long sequence of the factor approximants could be considered
\cite{Yukalov_13,Yukalov_18,Gluzman_19,Yukalov_20,Yukalov_21}. It has been shown that for 
finite values of the considered variable the accuracy of self-similar approximants is
comparable with that of heavy numerical calculations. In the present paper we concentrate 
on the most difficult and interesting challenge of finding strong-coupling limits of 
functions, especially their exponents, from the knowledge of only a few terms in their 
asymptotic weak-coupling expansions. This type of problems is difficult even for numerical 
methods. The approach is illustrated by several problems of quantum field theory.

\section{Convergence of self-similar factor approximants}

When a number of terms in a weak-coupling expansion are available and the asymptotic 
behavior in the strong-coupling limit is known, it is possible to study numerical 
convergence of the approximants. Below we illustrate this by several examples.    

\subsection{Zero-dimensional $\vp^4$ theory}

Let us start with the simple example of the so-called zero-dimensional $\vp^4$ 
theory characterized by the generating functional (partition function)
\be
\label{E1}
 Z(g) = \frac{1}{\sqrt{\pi}} \int_{-\infty}^\infty e^{-\vp^2 - g\vp^4} \; d\vp \; ,
\ee
with the coupling parameter $g \geq 0$. The weak-coupling asymptotic expansion reads 
as
\be
\label{E2}
Z_k(g) = \sum_{n=0}^k a_n g^n \qquad ( g \ra 0 ) \;   ,
\ee
where the coefficients are
$$
a_n = \frac{(-1)^n}{\sqrt{\pi}\; n!} \;\Gm\left( 2n + \frac{1}{2}\right) \; .
$$
Using only this weak-coupling expansion, we construct the self-similar factor 
approximants and study their strong-coupling limit, which gives
\be
\label{E3}
Z_k^*(g) \simeq B_k g^{\gm_k} \qquad ( g \ra \infty) \; .
\ee
The accuracy of the obtained strong-coupling exponents can be found by comparing the 
above limit with the known strong-coupling behavior
\be
\label{E4}
 Z(g) \simeq 1.022765 \; g^{-1/4} \qquad ( g\ra \infty) \; .
\ee
The results, shown in Table 1, demonstrate monotonic convergence to the exact limiting 
value $-0.25$. In the $16$-th order, we have
\be
\label{E5}
Z_{16}^*(g) \simeq 0.828 \; g^{-0.187} \qquad ( g\ra \infty) \;   .
\ee
 
\begin{table}[h!]
\centering
\caption{Strong-coupling exponents and their percentage errors for the generating 
functional of zero-dimensional $\vp^4$ theory.}
\vskip 3mm
\label{Table 1}
\renewcommand{\arraystretch}{1.25}
\begin{tabular}{|c|c|c|c|c|c|c|c|c|} \hline
$k$           &      2  &   4     &   6     &    8    &   10    &  12     &   14    &  16  \\ \hline
$\gm_k$       & $-0.09$ & $-0.13$ & $-0.15$ & $-0.16$ & $-0.17$ & $-0.18$ & $-0.18$ & $-0.19$ \\ \hline
$\ep(\gm_k)\%$& $-$63   & $-$48   & $-$41   & $-$36   & $-$32   & $-$29   & $-$27   & $-$25 \\ \hline
\end{tabular}
\end{table}

\subsection{One-dimensional anharmonic oscillator}

The one-dimensional anharmonic oscillator with the Hamiltonian
\be
\label{E6}
 H = - \; \frac{1}{2} \; \frac{d^2}{dx^2} +\frac{1}{2}\; x^2 + g x^4 \;  ,
\ee
where $g \geq 0$ and $-\infty<x< \infty$, imitates the one-dimensional $\vp^4$ theory.
The weak-coupling expansion of the ground-state energy is
\be
\label{E7}
E_k(g) \simeq \frac{1}{2} + \sum_{n=0}^k a_n g^n \qquad ( g \ra 0 )  \;  ,
\ee
with the coefficients that can be found in Refs. \cite{Bender_42,Hioe_43}. 

Constructing the factor approximants and looking for their strong-coupling limit 
\be
\label{E8}
E_k^*(g) \simeq B_k g^{\gm_k} \qquad ( g \ra \infty ) 
\ee
yields the results for the exponents shown in Table 2. The accuracy is found by the 
comparison with the known strong-coupling asymptotic behavior
\be
\label{E9}
 E(g) \simeq 0.667986 \; g^{1/3} \qquad ( g \ra \infty ) \;  .
\ee
In the $16$-th order, we get
\be
\label{E10}
 E_{16}^*(g) \simeq 0.736 \; g^{0.298} \qquad ( g \ra \infty ) \; .
\ee
Table 2 demonstrates monotonic numerical convergence. 

\begin{table}[h!]
\centering
\caption{Strong-coupling exponents and their percentage errors for the ground-state
energy of one-dimensional ahharmonic oscillator.}
\vskip 3mm
\label{Table 2}
\renewcommand{\arraystretch}{1.25}
\begin{tabular}{|c|c|c|c|c|c|c|c|c|} \hline
$k$           &     2  &   4   &   6    &    8   &   10   &  12    &   14   &  16  \\ \hline
$\gm_k$       &  0.18  & 0.23  &  0.26  &  0.27  &  0.28  &  0.29  &  0.29  &  0.30  \\ \hline
$\ep(\gm_k)\%$ & $-$47  & $-$31 & $-$23  & $-$18 & $-$16  & $-$13  & $-$12  & $-$11 \\ \hline
\end{tabular}
\end{table}

\subsection{Massive Schwinger model in lattice theory}    

One of the simplest nontrivial gauge-theory models is the Schwinger model 
\cite{Schwinger_44}. This is a lattice model of quantum electrodynamics in $1+1$
space-time dimensions. The model exhibits several phenomena typical of quantum
chromodynamics, such as confinement, chiral symmetry breaking with an axial anomaly, 
and a topological vacuum \cite{Banks_45,Carrol_46,Vary_47,Adam_48,Striganesh_49}.
The spectrum of excited states for a finite lattice, calculated by means of 
perturbation theory, is expressed through the series 
\be
\label{E11}
f_k(z) \simeq 1 + \sum_{k=1}^k a_n z^n \qquad ( z \ra 0 )
\ee
in powers of the variable $z \equiv 1/(ga)^4$, where $g$ is the coupling parameter and 
$a$ is the lattice spacing \cite{Hamer_50}. The coefficients for the vector boson are
$$
a_1 = 2 \; , \qquad a_2 = -10 \; , \qquad a_3 = 78.66667 \; , \qquad
a_4 = -7.362222\times 10^2 \; , 
$$
$$
a_5 = 7.572929\times 10^3 \; , \qquad a_6 = - 8.273669\times 10^4 \; , \qquad 
a_7 = 9.428034\times 10^5 \; , 
$$
$$
a_8 = -1.108358\times 10^7 \; , \qquad a_9 = 1.334636\times 10^8 \; , \qquad 
a_{10} = -1.637996 \times 10^9 \; .
$$   

Constructing factor approximants and considering their large-$z$ limit, we have
\be
\label{E12}  
f_k^*(z) \simeq B_k z^{\gm_k} \qquad ( z  \ra \infty ) \;  ,
\ee   
with the results for the large-$z$ exponent listed in Table 3. This is to be compared
with the known limiting behavior
\be
\label{E13}
 f(z) \simeq 1.1284 z^{1/4}  \qquad ( z \ra \infty) \; .   
\ee
For example, in the $10$-th order
\be
\label{E14}
f_{10}^*(z) \simeq 1.519 \; z^{0.2} \qquad ( z \ra \infty) \;   .
\ee

\begin{table}[h!]
\centering
\caption{Large-$z$ exponents and their percentage errors for the function $f(z)$ 
of the finite-lattice Schwinger model.}
\vskip 3mm
\label{Table 3}
\renewcommand{\arraystretch}{1.25}
\begin{tabular}{|c|c|c|c|c|c|} \hline
$k$           &     2   &   4   &   6     &    8   &   10     \\ \hline
$\gm_k$       &  0.167  & 0.185 &  0.193  &  0.198 &  0.200  \\ \hline
$\ep(\gm_k)\%$ &  $-$33  & $-$26 &  $-$23  &  $-$21 &  $-$20   \\ \hline
\end{tabular}
\end{table}

\subsection{Ground-state energy of Schwinger model}

The ground-state energy of the Schwinger model with a vector boson, in the continuum 
limit, can be found \cite{Carrol_46,Vary_47,Adam_48,Striganesh_49} as an expansion in 
powers of the dimensionless variable $x = m/g$, where $m$ is the electron mass and $g$ 
is the coupling parameter,
\be
\label{E15} 
\frac{E(x)}{g} \simeq 0.5642 - 0.219 x + 0.1907 x^2 \qquad ( x \ra 0 ) \; .
\ee

This short series allows us to construct only the second-order factor approximant
\be
\label{E16}
\frac{E_2^*(x)}{g}  = \frac{0.5642}{(1+1.35339 x)^{0.286805}} \;  .
\ee
In the large-$x$ limit this gives
\be
\label{E17}
\frac{E_2^*(x)}{g} \simeq 0.5173 x^{-0.287} \qquad ( x \ra \infty)  \; .
\ee
Comparing the large-$x$ exponent, with the known asymptotic behavior 
\cite{Striganesh_49,Hamer_50,Coleman_51,Hamer_52}
\be
\label{E18}
 \frac{E(x)}{g} \simeq 0.6417 x^{-1/3} \qquad ( x \ra \infty) \; ,
\ee
we find that the percentage error of the predicted exponent is $\ep(\gm_2)=-14\%$.

\subsection{Summary for considered examples} 

The above examples show that the knowledge of only a small-variable asymptotic expansion
makes it possible to extrapolate the small-variable expansion to finite-values of the
variable and even to predict the behavior of the corresponding function at asymptotically 
large values of the variable. When a number of terms in the small-variable series are known,
the sequence of the related self-similar approximants is shown to converge. The self-similar
extrapolation allows for sufficiently accurate evaluation of the large-variable exponent 
even when just a few terms of the small-variable expansion are available.

The reason why a small-variable expansion can be extrapolated to the finite and even 
infinite values of the variable lays in the following. The coefficients of the expansion
contain hidden information on the whole function which they are derived from. Separate 
coefficients do not allow for noticing this hidden information. However this information 
can be extracted by analyzing the relations between the coefficients. Self-similar 
approximation theory provides an instrument revealing the relations between the expansion 
coefficients and thus allowing for the reconstruction of the whole sought function.

\section{Exact reconstruction of Gell-Mann-Low functions}

Gell-Mann-Low functions in quantum field theory are usually calculated by means of loop 
expansions yielding series in powers of asymptotically small coupling parameters. However, 
the behaviour of these functions at strong coupling is of special interest. Below we 
consider the extrapolation of these functions to the arbitrary values of coupling parameters, 
including the limit to $\infty$, by employing self-similar factor approximants. A special 
attention will be paid to the study of the strong-coupling limit. In the present section, 
we demonstrate that in some cases, having just a few perturbative terms, self-similar 
approximants can reconstruct the sought Gell-Mann-Low function exactly.  

For this purpose, let us turn to the $N=1$ supersymmetric pure Yang-Mills theory whose 
exact beta function is known \cite{Novikov_52,Novikov_53,Shifman_54,Arkani_55,Arkani_56,Goriachuk}:
\be
\label{S1}
 \bt(g) = -\; \frac{3g^3 N_c}{16\pi^2(1-g^2N_c/8\pi^2)} \; .
\ee

If one resorts to perturbation theory with respect to the coupling $g$, one gets
\be
\label{S2}
 \bt_k(g) = - \; \frac{3g^3N_c}{16\pi^2} \sum_{n=0}^k b_n g^{2n} \;  ,
\ee
with the coefficients
$$
b_n = \left(\frac{N_c}{8\pi^2}\right)^n \; .
$$

For the second-order factor approximant, we have
\be
\label{S3}
\bt_2^*(g) = - \; \frac{3g^3N_c}{16\pi^2} \; \left( 1 + A_1 g^2 \right)^{n_1} \; .
\ee
Expanding Eq. (\ref{S3}) and comparing the expansion with series (\ref{S2}) yields
$$
n_1 = -1 \; , \qquad A_1 = - \; \frac{N_c}{8\pi^2} \; .
$$
Thus the second-order factor approximant results in the exact expression (\ref{S1}).
It is easy to check that all approximants of orders $k \geq 2$ give the exact beta 
function (\ref{S1}).

\section{Gell-Mann-Low function in $\vp^4$ field theory} 

Let us consider the $O(N)$ symmetric $\vp^4$ field theory. The Gell-Mann-Low function 
is defined as
\be
\label{13}
\bt(g) = \mu\; \frac{\prt g}{\prt\mu} \; ,
\ee
where $g=\lbd/(4\pi)^2$ is the coupling parameter and $\mu$ is renormalization scale.
This function has been found \cite{Kompaniets_21}, within minimal subtraction scheme, 
in the six-loop approximation: 
\be
\label{14}
\bt(g) \simeq g^2 \sum_{n=0}^k b_n g^n \qquad ( g \ra 0 ) \;  ,
\ee
with the coefficients
$$
b_0 = \frac{N+8}{3} \; , \qquad b_1 = -\; \frac{3N+14}{3} \; ,
$$
$$
b_2 = \frac{1}{216} \left[ 96(5N+22)\zeta(3) + 33 N^2 + 922 N + 2960 \right] \; ,
$$
$$
b_3 = -\; \frac{1}{3888} \left[ 1920 \left( 2N^2 + 55N + 186 \right) \zeta(5)
- 288(N+8)(5N+22)\zeta(4) + \right.
$$
$$
+ \left.
96\left( 63N^2 + 764 N + 2332 \right)\zeta(3) -
\left( 5N^3 - 6320 N^2 - 80456 N - 196648\right)  \right] \; ,
$$
$$
b_4 =  \frac{1}{62208} \left[  112896 \left( 14 N^2 + 189 N + 526\right)\zeta(7) -
768 \left( 6N^3 + 59N^2 - 446 N - 3264\right) \zeta^2(3) -  \right.
$$
$$
-
9600(N+8) \left( 2N^2 + 55N + 186\right) \zeta(6) +
256\left( 305 N^3 + 7466 N^2 + 66986 N + 165084\right)\zeta(5) -
$$
$$
-
288\left(63 N^3 + 1388 N^2 + 9532 N + 21120\right)\zeta(4) -
$$
$$
-
16\left( 9N^4 - 1248 N^3 - 67640 N^2 - 552280 N - 1314336\right)\zeta(3) +
$$
$$
+ \left.
13 N^4 + 12578 N^3 + 808496 N^2 + 6646336 N + 13177344 \right] \; ,
$$
$$
b_5 =  -\; \frac{1}{41990400} \left[  204800
\left( 1819N^3 + 97823 N^2 + 901051 N + 2150774\right)\zeta(9) +  \right.
$$
$$
+
14745600\left( N^3 + 65 N^2 + 619 N + 1502 \right)\zeta^3(3) +
$$
$$
+
995328 \left( 42N^3 + 2623 N^2 + 25074 N + 59984\right)\zeta(3,5) -
$$
$$
-
20736\left( 28882 N^3 + 820483 N^2 + 6403754 N + 14174864 \right) \zeta(8) -
$$
$$
-
5529600 \left( 8N^3 - 635 N^2 - 9150 N - 25944 \right) \zeta(3)\zeta(5) +
$$
$$
+
11520\left( 440N^4 + 126695 N^3 + 2181660 N^2 + 14313152 N + 29762136
\right)\zeta(7) +
$$
$$
+
207360 (N+8)\left( 6N^3 + 59N^2 - 446N - 3264\right)\zeta(3)\zeta(4) -
$$
$$
-
23040\left( 188 N^4 + 132 N^3 - 93363 N^2 - 862604 N - 2207484\right)\zeta^2(3)
-
$$
$$
-
28800\left( 595 N^4 + 20286 N^3 + 277914 N^2 + 1580792 N + 2998152\right)\zeta(6) +
$$
$$
+
5760\left( 4698 N^4 + 131827 N^3 + 2250906 N^2 + 14657556 N + 29409080\right)\zeta(5)
+
$$
$$
+
2160\left( 9 N^5 - 1176 N^4 - 88964 N^3 - 1283840N^2 - 6794096 N -
12473568\right)\zeta(4) -
$$
$$
- 720\left( 33 N^5 + 2970 N^4 - 477740 N^3 - 10084168 N^2 -
61017200 N - 117867424 \right)\zeta(3) -
$$
$$
- \left.
45\left( 29N^5 + 22644 N^4 - 3225892 N^3 - 88418816 N^2 - 536820560 N -
897712992\right)  \right] \; .
$$
Here $\zeta(3,5)$ denotes the double zeta function
$$
\zeta(3,5) = \sum_{1\leq n<m} \frac{1}{n^3m^5} = 0.037707673 \;   .
$$
The numerical values of the coefficients for the number of components from $N=0$ to 
$N=4$ are given in Table 4.

\begin{table}[h!]
\centering
\caption{Coefficients of weak-coupling expansion for the Gell-Man-Low function
of the $N$-component $\vp^4$ field theory, in the six-loop approximation.}
\vskip 3mm
\label{Table 4}
\renewcommand{\arraystretch}{1.25}
\begin{tabular}{|c|c|c|c|c|c|} \hline
$N$   &     0       &       1   &    2     &    3     &    4     \\ \hline
$b_0$ &    2.66667  &     3.0    &    3.33333 &    3.66667 &    4.0 \\ \hline
$b_1$ & $-$4.66667  & $-$5.66667 & $-$6.66667 & $-$7.66667 & $-$8.66667  \\ \hline
$b_2$ &  25.4571    &   32.5497  &  39.9478   &  47.6514   &   55.6606   \\ \hline
$b_3$ & $-$200.926  & $-$271.606 & $-$350.515 & $-$437.646 & $-$532.991  \\ \hline
$b_4$ &  2003.98    &  2848.57   &   3844.51  &   4998.62  &   6317.66     \\ \hline
$b_5$ &  $-$23314.7 & $-$34776.1 & $-$48999.1 & $-$66242.7 & $-$86768.4 \\ \hline
\end{tabular}
\end{table}

In the case of $N = 1$, the Gell-Mann-Low function is known in the seven-loop 
approximation \cite{Schnetz_22} having the coefficients
$$
b_0 = 3 \; , \qquad b_1 = -5.66667 \; , \qquad b_2 = 32.5497 \; , \qquad
b_3 = -271.606 \; ,
$$
$$
 b_4 = 2848.57 \; , \qquad b_5 = -34776.1 \; , \qquad
b_6 = 474651  \;   .
$$

We construct self-similar factor approximants for different $N$. Thus for $N=0$, 
we have  
$$
\bt_2^*(g) = \frac{2.66667 g^2}{( 1 + 9.16021 g)^{0.191044} } \qquad ( N = 0 ) \; ,
$$
\be
\label{15}
 \bt_4^*(g) = \frac{2.66667 g^2}{( 1 + 6.06377 g)^{0.194059}(1+15.5161 g)^{0.0369468} } \;  ,
\ee
which yields the strong-coupling limit
\be
\label{16}
 \bt_2^*(g) \simeq 1.747 g^{1.809} \; , \qquad  \bt_4^*(g) \simeq 1.699 g^{1.769}
\qquad ( N = 0 , ~ g \ra \infty) \; .
\ee

For $N = 2$, we find
$$
\bt_2^*(g) = \frac{3.33333 g^2}{( 1 + 9.98433 g)^{0.200314} } \qquad ( N = 2 ) \; ,
$$
\be
\label{17}
\bt_4^*(g) = \frac{3.33333 g^2}{( 1 + 5.79973 g)^{0.206354}(1+16.2195 g)^{0.0495207} } \;   ,
\ee
with the strong-coupling limit
\be
\label{18}
 \bt_2^*(g) \simeq 2.102 g^{1.7997} \; , \qquad  \bt_4^*(g) \simeq 2.020 g^{1.7441}
\qquad ( N = 2 , ~ g \ra \infty) \;  .
\ee

For $N = 3$, we obtain
$$
\bt_2^*(g) = \frac{3.66667 g^2}{( 1 + 10.3399 g)^{0.202218} } \qquad ( N = 3 ) \; ,
$$
\be
\label{19}
 \bt_4^*(g) = \frac{3.66667 g^2}{( 1 + 5.53502 g)^{0.212617}(1+16.526 g)^{0.055311} } \;  ,
\ee
giving the limit
\be
\label{20}
 \bt_2^*(g) \simeq 2.286 g^{1.798} \; , \qquad  \bt_4^*(g) \simeq 2.182 g^{1.732}
\qquad ( N = 3 , ~ g \ra \infty) \;   .
\ee

And for the case of $N = 4$, we get
$$
\bt_2^*(g) = \frac{4 g^2}{( 1 + 10.6781 g)^{0.202908} } \qquad ( N = 4 ) \; ,
$$
\be
\label{21}
 \bt_4^*(g) = \frac{4 g^2}{( 1 + 5.2767 g)^{0.219241}(1+16.8661 g)^{0.0598713} } \;  ,
\ee
with the strong-coupling limit
\be
\label{22}
 \bt_2^*(g) \simeq 2.474 g^{1.797} \; , \qquad  \bt_4^*(g) \simeq 2.345 g^{1.721}
\qquad ( N = 4 , ~ g \ra \infty) \;   .
\ee

For $N = 1$, we use the seven-loop expansion obtaining the functions
$$
\bt_2^*(g) = \frac{3 g^2}{( 1 + 9.59923 g)^{0.196775} } \qquad ( N = 1 ) \; ,
$$
$$
\bt_4^*(g) = \frac{3 g^2}{( 1 + 6.01378 g)^{0.200415}(1+15.9204 g)^{0.0429409} } \; ,
$$
\be
\label{23}
\bt_6^*(g) = 
\frac{3 g^2}{( 1 + 5.35692 g)^{0.186798}(1+13.7203 g)^{0.0596501}(1+22.0958 g)^{0.0031595} }\; ,
\ee
whose strong-coupling behavior is
$$
 \bt_2^*(g) \simeq 1.922 g^{1.803} \; , \qquad  
\bt_4^*(g) \simeq 1.859 g^{1.757} \; ,
$$
\be
\label{24}
\bt_6^*(g) \simeq 1.857 g^{1.750} \qquad ( N = 1 , ~ g \ra \infty) \;   .
\ee

Summarizing, we present in Table 5 the averaged results 
\be
\label{25}
B = \frac{1}{k} \sum_{n=1}^k B_{2n} \; , \qquad 
\gm = \frac{1}{k} \sum_{n=1}^k \gm_{2n}
\ee
for the amplitudes and exponents characterizing the strong-coupling limit
\be
\label{26}
 \bt(\gm) \simeq B g^\gm \qquad ( g \ra \infty ) \; ,
\ee
together with the dispersion between the subsequent approximants. 

The overall behaviour of the Gell-Mann-Low function of $\varphi^4$ field theory for
$N = 1$ is shown in Fig. 1, where the convergence of the approximants is evident. 
The behaviour of the Gell-Mann-Low functions for other $N$ is similar, only slightly 
differing from that for $N = 1$.      

In literature, it is possible to find the estimates for the strong-coupling exponent 
$\gamma$ in the case of $N = 1$. Thus Borel summation with conformal mapping gives 
$\gamma = 2$ \cite{Kazakov_23} or $\gamma = 1.9$ \cite{Chetyrkin_24}. A variational 
estimate \cite{Sissakian_25} yields $\gamma = 1.5$. Our result of $\gamma = 1.77$ for 
$N = 1$ is between those given by the Borel summation and variational calculations.  

\begin{table}[h!]
\centering
\caption{Strong-coupling amplitudes and exponents for the $N$-component
$\vp^4$ field theory, predicted by self-similar factor approximants.}
\vskip 3mm
\label{Table 5}
\renewcommand{\arraystretch}{1.25}
\begin{tabular}{|c|c|c|} \hline
$N$ &     B            &      $\gm$        \\ \hline
0   & $1.723 \pm 0.02$  &   $1.789 \pm 0.02$  \\ 
1   & $1.879 \pm 0.03$  &   $1.770 \pm 0.03$  \\ 
2   & $2.061 \pm 0.04$  &   $1.772 \pm 0.03$   \\ 
3   & $2.235 \pm 0.05$  &   $1.765 \pm 0.03$   \\
4   & $2.410 \pm 0.06$  &   $1.759 \pm 0.04$  \\ \hline
\end{tabular}
\end{table}

\section{Gell-Mann-Low function in quantum electrodynamics}

In quantum electrodynamics, the Gell-Mann-Low function in the renormalized minimal 
subtraction scheme $(\overline{\rm MS})$ reads as
\be
\label{27}
\bt(\al) = \mu^2 \; \frac{\prt}{\prt\mu^2} \left( \frac{\al}{\pi}\right) \; ,
\ee
where $\al$ is the renormalized $(\overline{\rm MS})$ scheme coupling parameter and 
$\mu$ is the $(\overline{\rm MS})$ scale parameter. The weak-coupling expansion in 
five-loop approximation, taking into account the electron, but neglecting the 
contributions of leptons with higher masses, that is, muons and tau-leptons, has 
the form \cite{Kataev_26}
\be
\label{28}
\bt(\al) \simeq  \left( \frac{\al}{\pi}\right)^2 
\sum_{n=0}^k b_n \left( \frac{\al}{\pi}\right)^n \;  ,
\ee
with the coefficients
$$
b_0 = \frac{1}{3} = 0.333333 \; , \qquad b_1 = \frac{1}{4} \; , \qquad
b_2 = -\; \frac{31}{288} = -0.107639 \; ,
$$
$$
 b_3 = -\; \frac{2785}{31104} \; - \; \frac{13}{36}\; \zeta(3) = - 0.523614\; , 
$$
$$
b_4 = -\; \frac{195067}{497664} \; - \; \frac{25}{96}\; \zeta(3) \; - \; 
\frac{13}{96}\; \zeta(4) \; + \; \frac{215}{96}\; \zeta(5) = 1.47072 \;   .
$$

From here, we find the factor approximants
$$
\bt_2^*(\al) = \frac{1}{3} \left( \frac{\al}{\pi}\right)^2  
\left( 1 + 1.61111 \; \frac{\al}{\pi} \right)^{0.465517} \;  , 
$$
\be
\label{29}
\bt_4^*(\al) = \frac{1}{3} \left( \frac{\al}{\pi}\right)^2  
\left( 1 + A_1 \; \frac{\al}{\pi} \right)^{n_1} 
\left( 1 + A_2 \; \frac{\al}{\pi} \right)^{n_2} \;
\ee
where
$$
A_1=1.394295 + 2.70199797\; i = A_2^* \; , \qquad
n_1=0.047762 - 0.11413974\; i = n_2^* \;   .
$$
Therefore in the strong-coupling limit, we have 
$$
\bt_2^*(\al) \simeq 0.4162 \left( \frac{\al}{\pi} \right)^{2.4655} \; ,
$$
\be
\label{30}
\bt_4^*(\al) \simeq 0.4759 \left( \frac{\al}{\pi} \right)^{2.0955} \qquad
\left( \frac{\al}{\pi}  \ra \infty \right) \;  .
\ee
Defining average quantities, we see that the strong-coupling behaviour of the 
Gell-Mann-Low function
\be
\label{31}
\bt(\al) \simeq B \left( \frac{\al}{\pi}\right)^\gm \qquad
\left( \frac{\al}{\pi} \ra \infty \right)
\ee
can be characterized by the amplitude and exponent
\be
\label{32}
 B = 0.446 \pm 0.03 \; , \qquad \gm = 2.281 \pm 0.19 \;  .
\ee

The dependence of the Gell-Mann-Low function for quantum electrodynamics is presented
in Fig. 2. Its overall behaviour only slightly depends on the chosen scheme. Thus, 
accepting the coefficients of the weak-coupling expansion found in the on-shell scheme 
or in the momentum subtraction scheme \cite{Kataev_26} results in $B=0.433,\;\gm=2.189$ 
in the on-shell scheme and $B=0.493,\;\gm=2.309$ in the momentum subtraction scheme.  
 
The running coupling defined by equation (\ref{27}), with the beta function represented 
by a factor approximant, increases from zero to infinity. As the boundary condition, we 
can take the value $\alpha(m_Z)=0.007815$ at the $Z$-boson mass $m_Z=91.1876$ GeV. 
Then the logarithmic divergence occurs at $\mu_0=8.584\times 10^{260}$ GeV, where
$$
\al \simeq \frac{2.743}{(\ln(\mu_0/\mu))^{0.682}} \qquad ( \mu \ra \mu_0 - 0) \; .
$$ 
The value of $\mu_0$ is much larger than the point of the simple Landau pole that is 
of the order of $10^{30}-10^{40}$ GeV \cite{Deur_27}. The value of $\mu_0$ is so large 
that practically can be considered as infinity.

\section{Gell-Mann-Low function in quantum chromodynamics} 

The Gell-Mann-Low function in quantum chromodynamics is defined by the equation
\be
\label{33}
\bt(\al_s) = \mu^2 \; \frac{\prt a_s}{\prt\mu^2} \qquad
\left( a_s \equiv \frac{\al_s}{\pi} \right) \; ,
\ee
where $\alpha_s$ is the quark-gluon coupling and $\mu$ is the normalization scale. We 
keep in mind the realistic case of three colours $(N_c = 3)$. In the five-loop 
approximation, the weak-coupling expansion is
\be
\label{34}
 \bt(\al_s) \simeq - a_s^2 \sum_{n=0}^k b_n a_s^n \qquad ( a_s \ra 0 ) \;  .
\ee
Within the minimal subtraction scheme $(\overline{\rm MS})$, the coefficients are
\cite{Luthe_27,Baikov_28,Herzog_29}
$$
b_0 = 2.75 - 0.166667\; n_f \; , \qquad b_1 = 6.375 - 0.791667\; n_f \; ,
$$
$$
b_2 = 22.3203 - 4.36892\; n_f + 0.0940394 \; n_f^2 \; ,
$$
$$
b_3 = 114.23 - 27.1339\; n_f + 1.58238 \; n_f^2 + 0.0058567 \; n_f^3\; ,
$$
$$
b_4 = 524.56 - 181.8\; n_f + 17.16 \; n_f^2 - 0.22586 \; n_f^3 -
0.0017993 \; n_f^4 \;  ,
$$   
with $n_f$ being the number of quark flavors. 

It turns out that factor approximants as real functions exist not for all $n_f$. 
But they do exist for the physically realistic number of flavors $n_f = 6$. For this 
case, the available factor approximants are
$$
\bt_2^*(\al_s) = -1.75 a_s^2 ( 1 + 1.5536 a_s)^{0.597706} \; , 
$$
\be
\label{35}
\bt_3^*(\al_s) = -1.75 a_s^2 ( 1 + a_s)^{0.91227} ( 1 + 32.5316 a_s)^{0.0005011}\;  .
\ee
This gives the strong-coupling limit 
\be
\label{36}
 \bt_2^*(\al_s) \simeq - 2.2772 a_s^{2.5977} \; , \qquad
\bt_3^*(\al_s) \simeq - 1.7531 a_s^{2.9128} \qquad ( a_s \ra \infty ) \; .
\ee
In that way, the strong-coupling limit of the Gell-Mann-Low function
\be
\label{37}
\bt(\al_s) \simeq - B a_s^\gm \qquad ( a_s \ra \infty )
\ee
is characterized by the amplitude and exponent
\be
\label{38}
B = -2.015 \pm 0.26 \; , \qquad \gm = 2.755 \pm 0.16 \;  .
\ee
We are not aware of other reliable estimates of these characteristics that could 
be compared with our result for the physically interesting case of the flavour number 
$n_f=6$. The behaviour of the Gell-Mann-Low function of quantum chromodynamics for
this flavour number, as a function of the coupling parameter, is shown in Fig. 3. 
For varying $n_f$, the Banks-Zaks \cite{Banks_30} fixed point exists in the region 
$8.05 \leq n_f \leq 16.5$.  
 
The running coupling, defined by Eq. (\ref{33}), with the boundary condition
$\alpha_s(m_Z) = 0.1181$ logarithmically grows when $\mu$ tends to $\mu_c = 0.1$ GeV
from above,
$$
\al_s \simeq \frac{0.907}{(\ln(\mu/\mu_c))^{0.626}} \qquad 
( \mu \ra \mu_c + 0) \;  .
$$  

The appearance of a pole at $\mu_c$ characterizes the scale at which perturbative QCD 
breaks down, so that the series (\ref{34}) as such, and hence their extrapolation, become 
invalid. The value of $\mu_c$ can be associated with the confinement scale, or equivalently 
the hadronic mass scale and nonperturbative effects, such as the arising bound states 
\cite{Deur_27}. The smaller the value of $\mu_c$, i.e., the smaller the momentum scale at 
which the divergence occurs, the slower the increase of $\alpha_s(\mu)$ as $\mu$ decreases. 
This would imply the effective extrapolation of perturbative expressions to smaller 
momentum scales \cite{Deur_27}. The value of $\mu_c$ is really much smaller than the point 
of the Landau pole which in the $(\overline{\rm MS})$ scheme for $n_f = 6$ happens at the 
point $0.9$ GeV \cite{Bethke_31}.

\section{Discussion}

In the above examples, we have considered the cases where the large-variable behavior 
is of power-law. As is demonstrated, for these cases, self-similar factor approximants 
can provide good extrapolation of small-variable Taylor-type asymptotic expansions to 
the range of finite variables and even for the large-variable limit. Moreover, in some 
cases these approximants, using only a small-variable expansion, are able to reconstruct 
the sought function exactly, as in the case of the Gell-Mann-Low function for the 
supersymmetric pure Yang-Mills theory and in some other cases to be considered below.  
    
Two related natural questions arise: How well the self-similar factor approximants could 
extrapolate the functions with the large-variable behavior different from the power-law, 
such as exponential and logarithmic behavior? And the other question is: What would be 
other examples of the exact function reconstruction by means of self-similar factor 
approximants?

\subsection{Class of exactly reproducible functions}

First of all, let us notice that there exists a class of real-valued functions exactly 
reproducible by factor approximants. These are the functions having the form 
\be
\label{39}
  R_{k_M}(x) = \prod_{i=1}^M \; P_{m_i}^{\al_i}(x) 
\ee
of the product of polynomials
$$
 P_{m_i}(x) = c_{i0} + c_{i1} x + c_{i2} x^2 + \cdots + 
c_{im_i} x^{m_i} \;   , 
$$  
where $m_i$ are integers; the powers $\alpha_i$ and coefficients $c_{ij}$ can be 
complex-valued numbers entering $R_{k_M}(x)$ in complex conjugate pairs so that 
$R_{k_M}(x)$ be real, and
$$
k_M = \sum_{i=1}^M m_i + M   .
$$
This follows from the fact that a polynomial $P_{m_i}(x)$ can be represented as
$$
 P_{m_i}(x) = c_{i0} \prod_{j=1}^{m_i}\; ( 1 + b_{ij} x ) \; ,
$$
with $b_{ij}$ being expressed through $c_{ij}$. Then the function $R_{k_M}(x)$ can be 
reduced to the form
\be
\label{40}
 R_{k_M}(x) = \prod_{i=1}^M \; c_{i0} \; 
\prod_{j=1}^{m_i} \; ( 1 + b_{ij} x )^{\al_i} \;  ,
\ee
which is nothing but a particular case of a factor approximant possessing the same 
asymptotic expansion as the given function (\ref{39}). 
   
If all powers $\alpha_i$ were $\pm 1$, then the function $R_k(x)$ would reduce to a Pad\'e
approximant that is a rational function. However the powers $\alpha_i$ are not necessarily 
integers. Hence factor approximants also include irrational functions that can be reproduced
exactly.

\subsection{Exact reconstruction of exponential functions}

Moreover, factor approximants can exactly reproduce transcendental functions, such as the 
exponential function $\exp(x)$, where $x$ can take any complex value. Let us consider the
standard $k$-order expansion of the exponential function 
\be
\label{41}
 e_k(x) = \sum_{n=0}^k \frac{x^n}{n!} \; .
\ee
The second-order factor approximant is
$$
e_2^*(x) = (1 + Ax)^n \;  .
$$
Expanding this in powers of $x$ and comparing with $e_k(x)$ yields the equations
$$
An = 1 \; , \qquad A^2 n (n-1) = 1 \; .
$$
The sole solution to these equations is $A=1/n$ with $n\ra\infty$. Therefore already 
the second-order factor approximant gives exactly the exponential function
\be
\label{42}
e_2^*(x) = \lim_{n\ra\infty}\; \left( 1 + \frac{1}{n}\; x \right)^n = e^x \;   .
\ee
It is easy to check that all factor approximants of the order $k \geq 2$ reconstruct 
the exponential function exactly.

\subsection{Exact solution of nonlinear equations}

Some nonlinear differential equations can be solved exactly by looking for solutions 
in the form of asymptotic series and then constructing factor approximants. For instance, 
let us consider the nonlinear singular problem
\be
\label{43}
 (\ep y + t ) \; \frac{dy}{dt} + y - 1 = 0   
\ee
with the initial condition $y(0)=2$. This kind of equation is met in different 
applications \cite{Nayfeh_75,Hinch_76}. It is called singular since it does not allow 
for the use of perturbation theory in powers of the parameter $\varepsilon$. 

The parameter $\ep$ can be hidden in the renotation
$$
 t = \ep x \; , \qquad y = z - x  
$$
resulting in the equation
$$
 z \; \frac{dz}{dt} - x - 1 = 0 \;  ,
$$
with the initial condition $z(0) = 2$. Looking for the solution at asymptotically small 
$x$ implies the consideration of the series
$$
z_k(x) =  \sum_{n=0}^k a_n x^n \; ,
$$ 
whose coefficients can be found by substituting these series into the equation. Thus
$a_0 = 2$, $a_1 = 1/2$, $a_3 = 3/16$, and so on. Constructing the fourth-order factor 
approximant gives
$$
 z_4^*(x) = 2 (1 + A_1x)^{n_1} (1 + A_2 x)^{n_2} \;  ,
$$
with the parameters
$$
A_1 = \frac{1}{4} \; ( 1 - i\sqrt{3} ) = A_2^* \; , \qquad 
n_1 = n_2 = \frac{1}{2} \;   .
$$
Returning to the initial variables yields the function
\be
\label{44}
 y_4^*(t) = \sqrt{4 + \frac{2t}{\ep} + \frac{t^2}{\ep^2} } \; - \; 
\frac{t}{\ep}  
\ee
that is the exact solution of the given equation. The same exact solution results 
for any factor approximant of $k \geq 4$.  

Exact solutions of several other nonlnear differential equations can also be found 
by employing the summation of asymptotic series by means of the self-similar factor 
approximants \cite{Yukalova_74}.

\subsection{Exponential behavior: Bose-Einstein distribution}

As is shown above, the purely exponential behavior is reproduced by factor approximants 
exactly. Then we should expect that a behavior close to the purely exponential could be 
well approximated by these approximants. Let us consider the well known Bose-Einstein 
distribution
\be
\label{45}
 f_B(x) = \frac{1}{e^x-1} \; .
\ee
Suppose, only the expansion at small $x$, 
\be
\label{46}
 f_B(x) \simeq \frac{1}{x} \; - \; \frac{1}{2} \; + \; \frac{x}{12} \; - \;
\frac{x^3}{720} \; + \; \frac{x^5}{30240} \; -\; \frac{x^7}{1209600} \; + \;
\ldots \qquad ( x \ra 0) \; ,
\ee 
is available for us, and we do not know what function it represents. 

In the standard way, we construct factor approximants $f_k^*(x)$ that extrapolate 
expansion (\ref{46}) from asymptotically small $x$ to finite values of the latter. Our 
major interest is in the approximants providing the extrapolation to the large variable 
behavior with respect to $x\ra\infty$. By their structure, the factor approximants give
the power-law behavior at infinity, for instance
$$
 f_4^*(x) \simeq 2.94\cdot \frac{10^{-3}}{x^3} \; , \qquad
f_8^*(x) \simeq  1.1 \cdot \frac{10^{-7}}{x^5} \; , \qquad 
f_{12}^*(x) \simeq 2.97\cdot \frac{10^{-14}}{x^7} \qquad (x\ra\infty) \; .
$$  
As we see, these values quickly diminish, telling us that the real behavior at large 
$x$ is faster then of power law, probably, of exponential type. If we expect that the 
large-variable behavior is exponential, we can employ another variant of self-similar 
approximants, i.e. the self-similar exponential approximants \cite{Yukalov_77}.
However, since, as is proved above, the factor approximants well approximate the 
exponential behavior, they should provide rather good accuracy for the extrapolation 
of the distribution (45), which is demonstrated in Fig. 4. The exact Bose-Einstein 
distribution (\ref{45}) and its factor approximants practically coincide in a large 
range of $x$, because of which in Fig. 4 we show their difference.

\subsection{Exponential behavior: Fermi-Dirac distribution}

Similarly, we can consider the Fermi-Dirac distribution
\be
\label{47}
 f_F(x) = \frac{1}{e^x+1} \;  .
\ee
Again assume that we possess only the small-variable expansion
\be
\label{48}
 f_F(x) \simeq \frac{1}{2} \; - \; \frac{x}{4} \; + \; \frac{x^3}{48} \; - \;
\frac{x^5}{480} \; + \; \frac{17x^7}{80 640} \; -\; \frac{31x^9}{1 451 520} \; + \;
\ldots \qquad ( x \ra 0 )
\ee
and are not aware of the function it corresponds to. Factor approximants $f_k^*(x)$ 
that extrapolate the series (\ref{48}) to the large-variable range again quickly 
diminish, for instance
$$
f_4^*(x) \simeq 6.7 \cdot \frac{10^{-2}}{x^2} \; , \qquad 
f_8^*(x) \simeq 3.4 \cdot \frac{10^{-5}}{x^4} \; \qquad
f_{12}^*(x) \simeq 1.05 \cdot \frac{10^{-10}}{x^6} \qquad (x\ra\infty) \;  .
$$ 
This hints that the large-variable behavior should be faster than of power law. 
Nevertheless, since factor approximants well approximate the purely exponential behavior,
they should provide quite accurate approximation for the considered distribution. Again,
the exact distribution (\ref{47}) and its factor approximants practically coincide in a 
large region of $x$, because of which in Fig. 5 the related differences are shown.

\subsection{Logarithmic behavior: no additional information}

Now let us turn to functions with logarithmic asymptotic behavior at large values of 
a variable. Let us take the function
\be
\label{49}
 f(x) =1 + \ln\; \left( \frac{1+\sqrt{1+x}}{2} \right) \;  ,
\ee
with the logarithmic asymptotic behavior at large $x$, 
\be
\label{50}
 f(x) \simeq \frac{1}{2} \ln \; x \qquad ( x\ra \infty ) \; .
\ee

Let us pretend that we know neither the function itself nor its behavior at large $x$,
but what available is only the asymptotic expansion at small $x$,
\be
\label{51}
f(x) \simeq 1 \; + \; \frac{1}{4} \; x \; - \; \frac{3}{32} \; x^2 \; + \;
\frac{5}{96}\; x^3 \; - \; \frac{35}{1024}\; x^4 \; + \; 
\frac{63}{2560}\;  x^5 \; - \; \frac{77}{4096}\;  x^6 \; +
\ldots \qquad ( x \ra 0 ) \; .
\ee
Constructing factor approximants from these series, we find that their large-variable
behavior demonstrates not so fast variation of the amplitude and power:
$$
f_2^*(x) \simeq x^{0.25} \; , \qquad f_4^*(x) \simeq 1.195\; x^{0.182} \; ,
\qquad f_6^*(x) \simeq 1.137 \; x^{0.197} \; ,
$$
$$
f_8^*(x) \simeq 1.237\; x^{0.1747} \; , \qquad 
f_{10}^*(x) \simeq 1.235\; x^{0.1749} \; , \qquad
f_{12}^*(x) \simeq 1.310\; x^{0.162} \qquad ( x \ra \infty) \;  .
$$
Although, as is typical of the factor approximants, the large-variable dependence is 
of power law, the factor approximants provide reasonable accuracy in a wide region of 
the variable $x$, as is seen in Fig. 6(a).

\subsection{Logarithmic behavior: known character of large-variable limit}

A different situation develops when, although the function itself is not known, but 
there is information that the large-variable behavior is expected to be logarithmic. 
Then it is reasonable to deal not with series (\ref{51}), but with the exponential of 
it. This implies that a finite series $f_k(x)$, corresponding to the truncated series 
(\ref{51}), is exponentiated considering
\be
\label{52}
g_k(x) = \exp\{ f_k(x) \}   
\ee
and then expanding the latter up to the $k$-th order. In the case of series (\ref{51}), 
we have 
\be
\label{53}
g(x) \simeq e \left( 1 + \frac{1}{4}\; x \; - \; \frac{1}{16}\; x^2 \; + \;
\frac{1}{32}\; x^3 \; - \; \frac{5}{256}\; x^4 \; + \;
\frac{7}{512} \;   x^5 \; - \;
\frac{21}{2048} \;  x^6 \; + \; \ldots \; \right) \quad
( x \ra 0 )  \;  .
\ee
The latter series is used for constructing the factor approximants $g_k^*(x)$, after 
which, inverting transformation (\ref{52}), we return to the expressions
\be
\label{54} 
 F_k^*(x) = \ln \; g_k^*(x)  
\ee
approximating the sought function. Here we denote the final approximants as $F_k^*(x)$ 
to distinguish them from the approximants $f_k^*(x)$ obtained directly from series (\ref{51}). 
 
The large-variable behavior of the approximants $F_k^*(x)$ is of correct logarithmic 
form, for example
$$
F_2^*(x) \simeq 0.333\ln x \; , \qquad  F_4^*(x) \simeq 0.4\ln x \; , 
\qquad F_6^*(x) \simeq 0.429\ln x \; ,
$$
$$
F_8^*(x) \simeq 0.444\ln x \; , \qquad  F_{10}^*(x) \simeq 0.456\ln x \; , 
\qquad F_{12}^*(x) \simeq 0.462\ln x \qquad ( x \ra \infty) \;  .
$$
The coefficient at the logarithm converges to the exact value $0.5$, in agreement 
with limit (\ref{50}). The overall behavior of the approximants $F_k^*(x)$ is shown 
in Fig. 6(b). As it should be expected, the additional information on the behavior of 
the function essentially improves the accuracy of the approximation.

\section{Conclusion}

We have suggested a method, based on self-similar approximation theory, allowing for 
the extrapolation of expressions from weak-coupling asymptotic expansions to the region 
of arbitrary values of coupling parameters, including their asymptotically large values. 
The region of large coupling parameters is of special interest because of its physical 
importance and because mathematically this region is the most difficult for the 
extrapolation that uses only the coefficients of weak-coupling expansions. 

In those cases, where a number of terms in the small-variable expansion are available,
the method is shown to posses numerical convergence. Good accuracy can be obtained even
for the expansions with a few terms. The extrapolation of perturbative series for the 
Gell-Mann-Low functions of the $O(N)$ symmetric $\varphi^4$ field theory, quantum 
electrodynamics, and quantum chromodynamics is demonstrated. 

In some cases, the method can transform perturbative series to the {\it exact} expression 
valid for arbitrary values of the variable. Such an exact reconstruction is illustrated 
for the Gell-Mann-Low function of a supersymmetric pure Yang-Mills theory and for several 
other examples.  

By their construction, self-similar factor approximants, for the variable tending to 
infinity, give a power-law behavior. The possibility is discussed of extrapolating 
functions with different types of the large-variable limits, including not only power-law 
behavior, but also logarithmic and exponential behavior. The purely exponential function
is shown to be reconstructed exactly by the factor approximants of any order starting 
from second. More complicated functions with exponential or logarithmic large-variable 
behavior can be well approximated in a wide range of the variable. If the additional 
information on the character of the large-variable limit is available, the accuracy of
the approximants can be essentially improved.

\newpage

\begin{center}
{\Large{\bf Figure Captions}}
\end{center}

\vskip 2cm

{\bf Figure 1}. Gell-Mann-Low function of $\varphi^4$ $O(1)$ symmetric field theory as 
a function of the coupling parameter $g$. The convergence of the approximants of second, 
fourth, and sixth order is evident. 

\vskip 1cm
{\bf Figure 2}. Gell-Mann-Low function of quantum electrodynamics as a function of the 
coupling parameter $\alpha$. 

\vskip 1cm
{\bf Figure 3}. Gell-Mann-Low function of quantum chromodynamics for the flavour number 
$n_f=6$ as a function of the coupling parameter $\al_s$. 

\vskip 1cm
{\bf Figure 4}. Difference between the Bose-Einstein distribution (\ref{45}) and its 
self-similar factor approximants: (a) $f_B(x) - f_4^*(x)$; (b) $f_B(x) - f_8^*(x)$;
(c) $f_B(x) - f_{12}^*(x)$. 

\vskip 1cm
{\bf Figure 5}. Difference between the Fermi-Dirac distribution (\ref{47}) and its
self-similar factor approximants: (a) $f_F(x) - f_4^*(x)$; (b) $f_F(x) - f_8^*(x)$;
(c) $f_F(x) - f_{12}^*(x)$. 

\vskip 1cm
{\bf Figure 6}. (a) Function (\ref{49}) (solid line) and its self-similar factor 
approximants obtained from the direct series (\ref{51}): $f_4^*(x)$ (dotted line), 
$f_8^*(x)$ (dash-dotted line), and $f_{12}^*(x)$ (dashed line). (b) Function (\ref{49}) 
(solid line) and its self-similar factor approximants obtained from expansion (\ref{53}):
$F_4^*(x)$ (dotted line), $F_8^*(x)$ (dash-dotted line), and $F_{12}^*(x)$ (dashed line).

\newpage

\begin{figure}[ht]
\centerline{
\includegraphics[width=10cm]{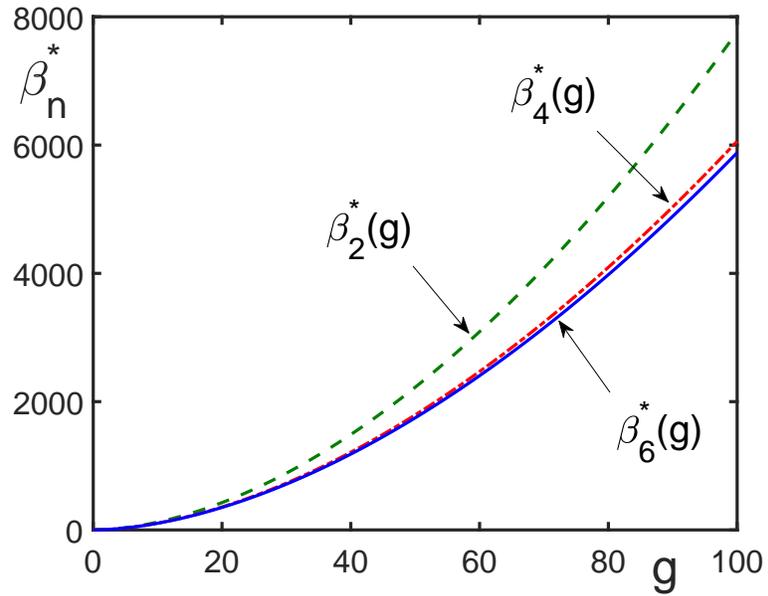} }
\vskip 3mm
\caption{Gell-Mann-Low function of $\vp^4$ $O(1)$ symmetric field theory as a function
of the coupling parameter $g$. The convergence of the approximants of second, fourth, 
and sixth order is evident.}
\label{fig:Fig.1}
\end{figure}

\newpage

\begin{figure}[ht]
\centerline{
\includegraphics[width=10cm]{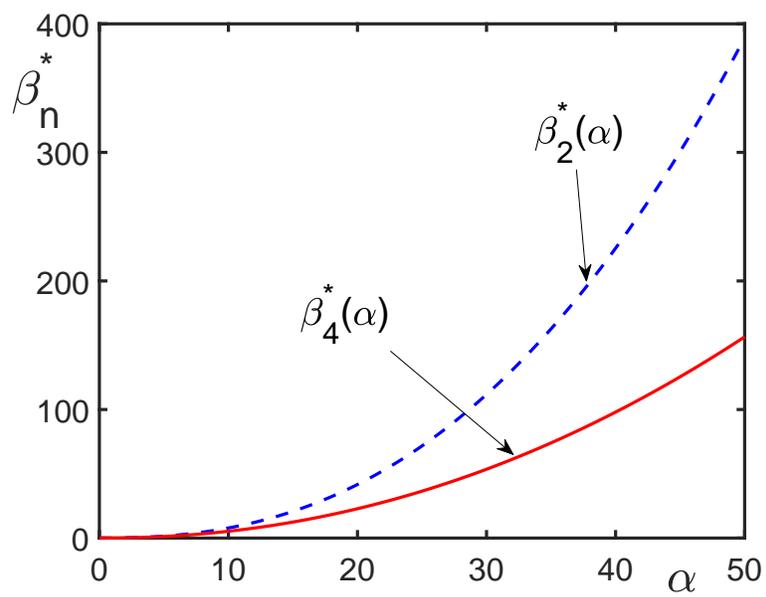} }
\vskip 3mm
\caption{Gell-Mann-Low function of quantum electrodynamics as a function of the coupling
parameter $\al$.}
\label{fig:Fig.2}
\end{figure}

\newpage

\begin{figure}[ht]
\centerline{
\includegraphics[width=10cm]{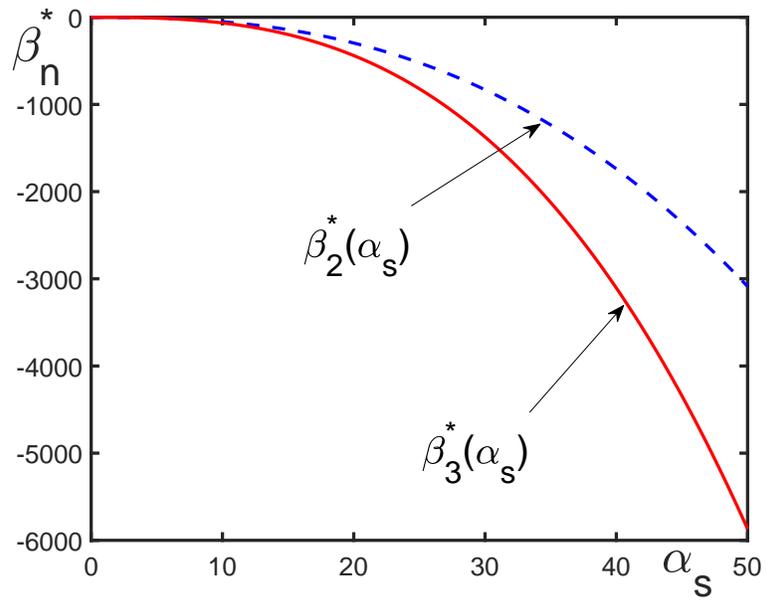} }
\vskip 3mm
\caption{Gell-Mann-Low function of quantum chromodynamics for the flavour number $n_f=6$ 
as a function of the coupling parameter $\al_s$.}
\label{fig:Fig.3}
\end{figure}

\newpage

\begin{figure}[ht]
\centerline{
\hbox{ \includegraphics[width=8cm]{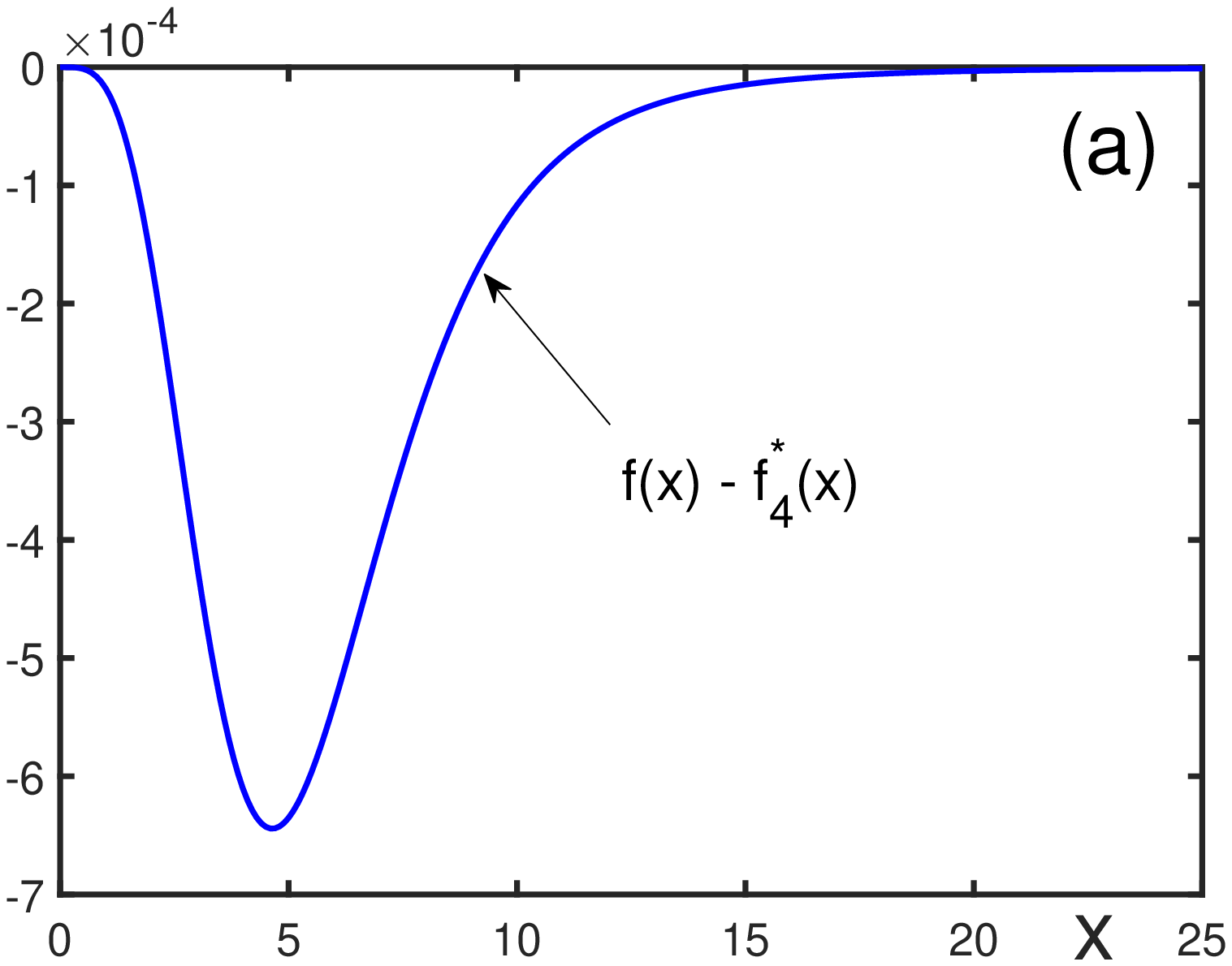} \hspace{1cm}
\includegraphics[width=8cm]{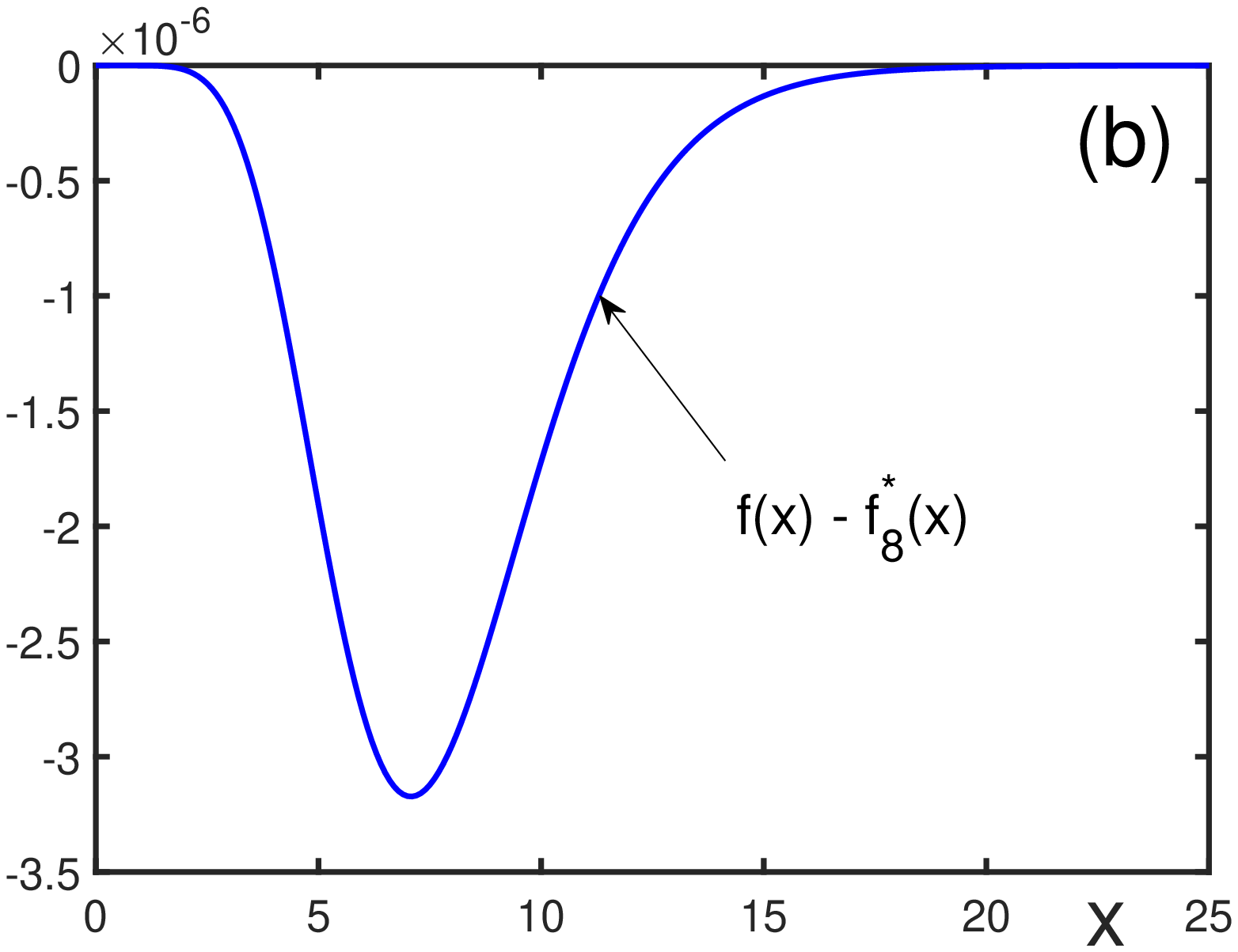}  } }
\vspace{12pt}
\centerline{
\hbox{ \includegraphics[width=8cm]{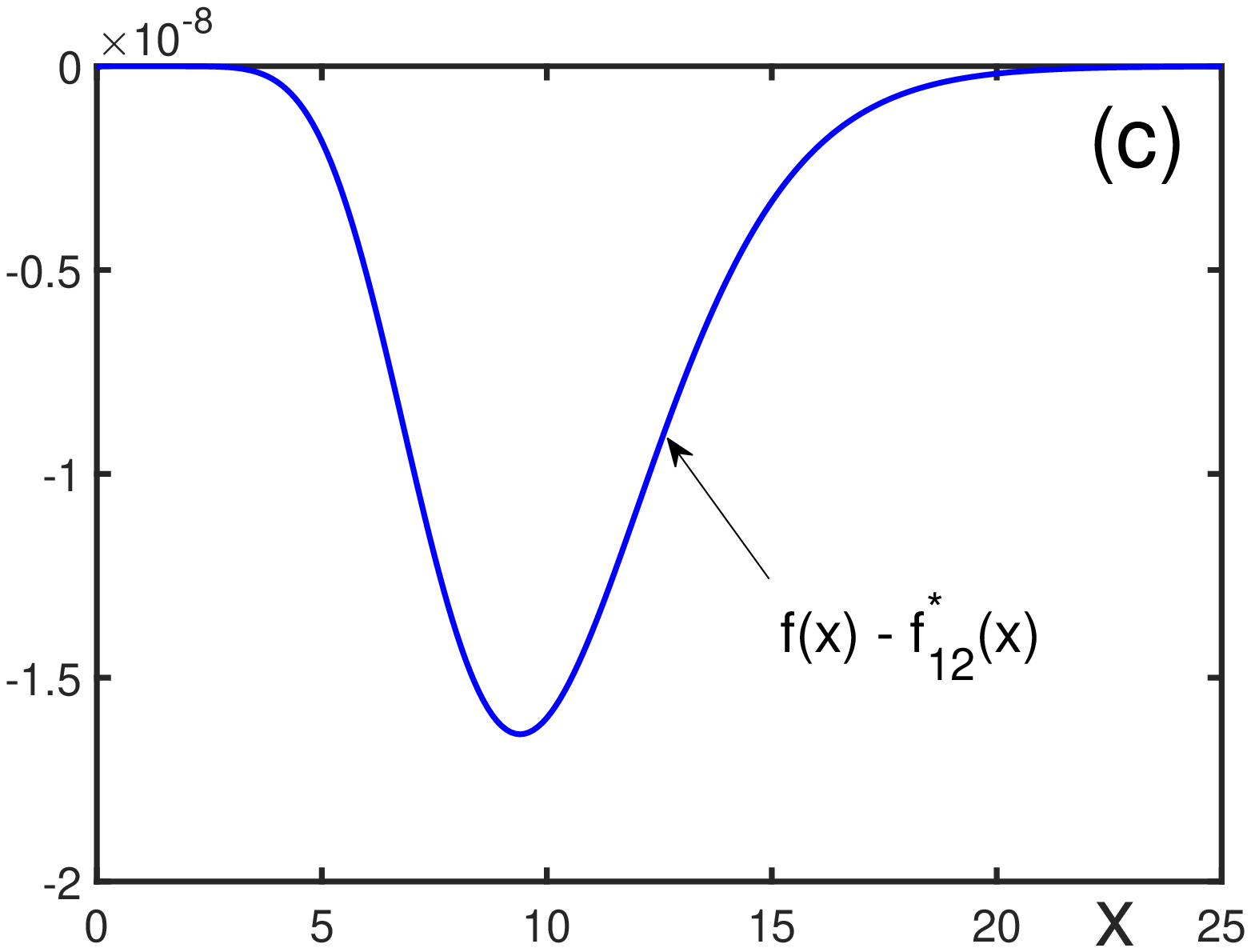} } }
\vskip 3mm
\caption{Difference between the Bose-Einstein distribution (\ref{45}) and its 
self-similar factor approximants: (a) $f_B(x) - f_4^*(x)$; (b) $f_B(x) - f_8^*(x)$;
(c) $f_B(x) - f_{12}^*(x)$.
}
\label{fig:Fig.4}
\end{figure}

\newpage

\begin{figure}[ht]
\centerline{
\hbox{ \includegraphics[width=8cm]{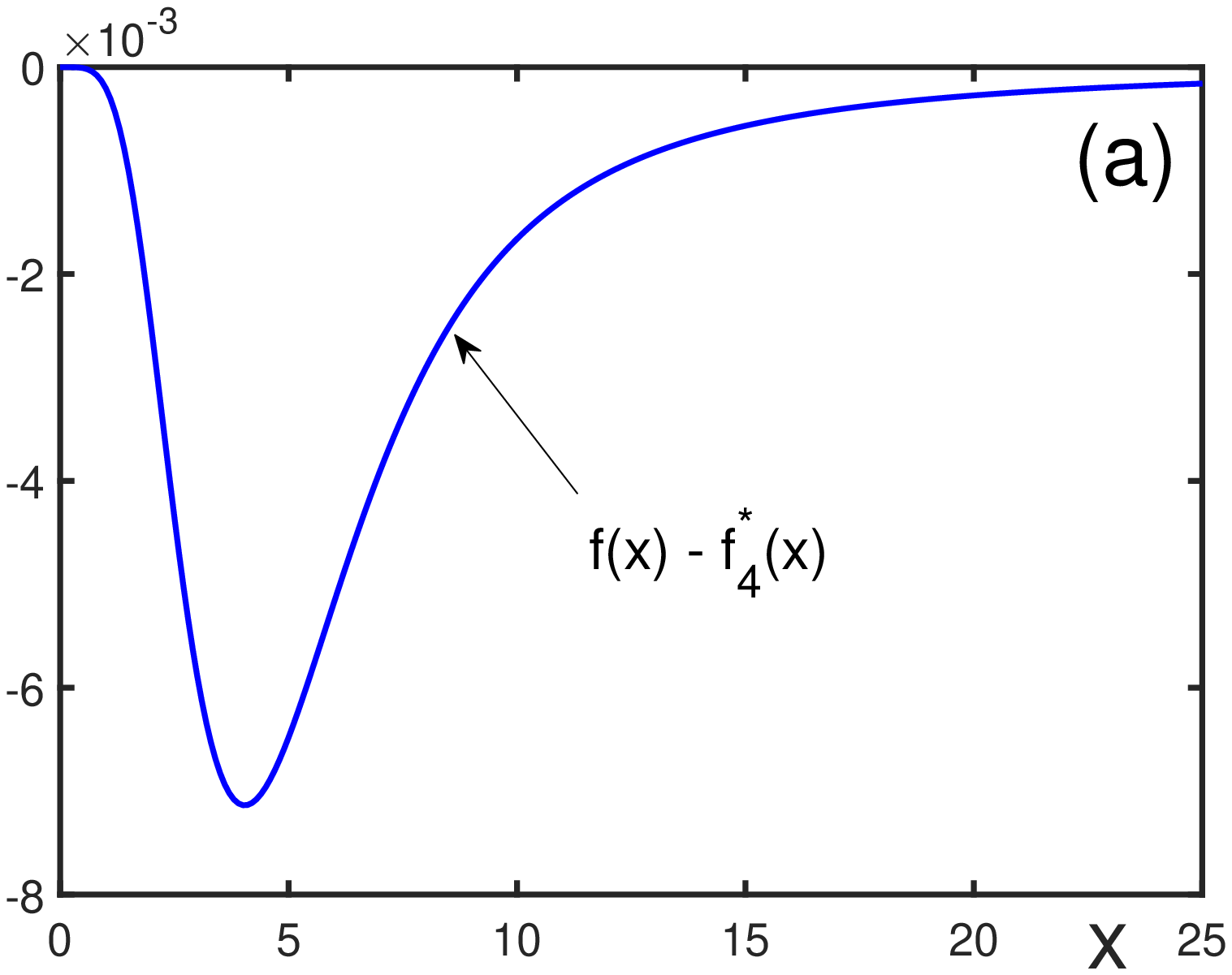} \hspace{1cm}
\includegraphics[width=8cm]{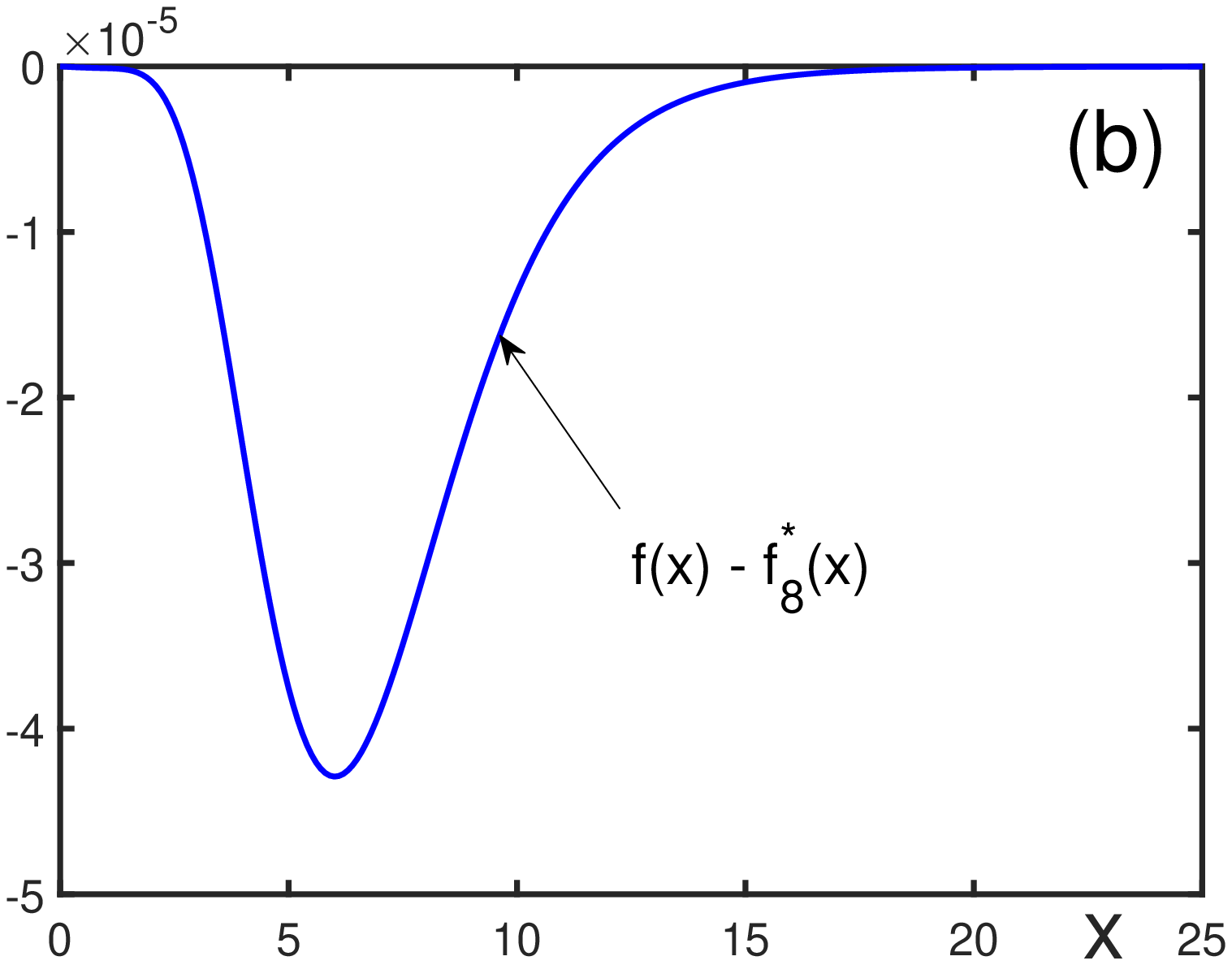}  } }
\vspace{12pt}
\centerline{
\hbox{ \includegraphics[width=8cm]{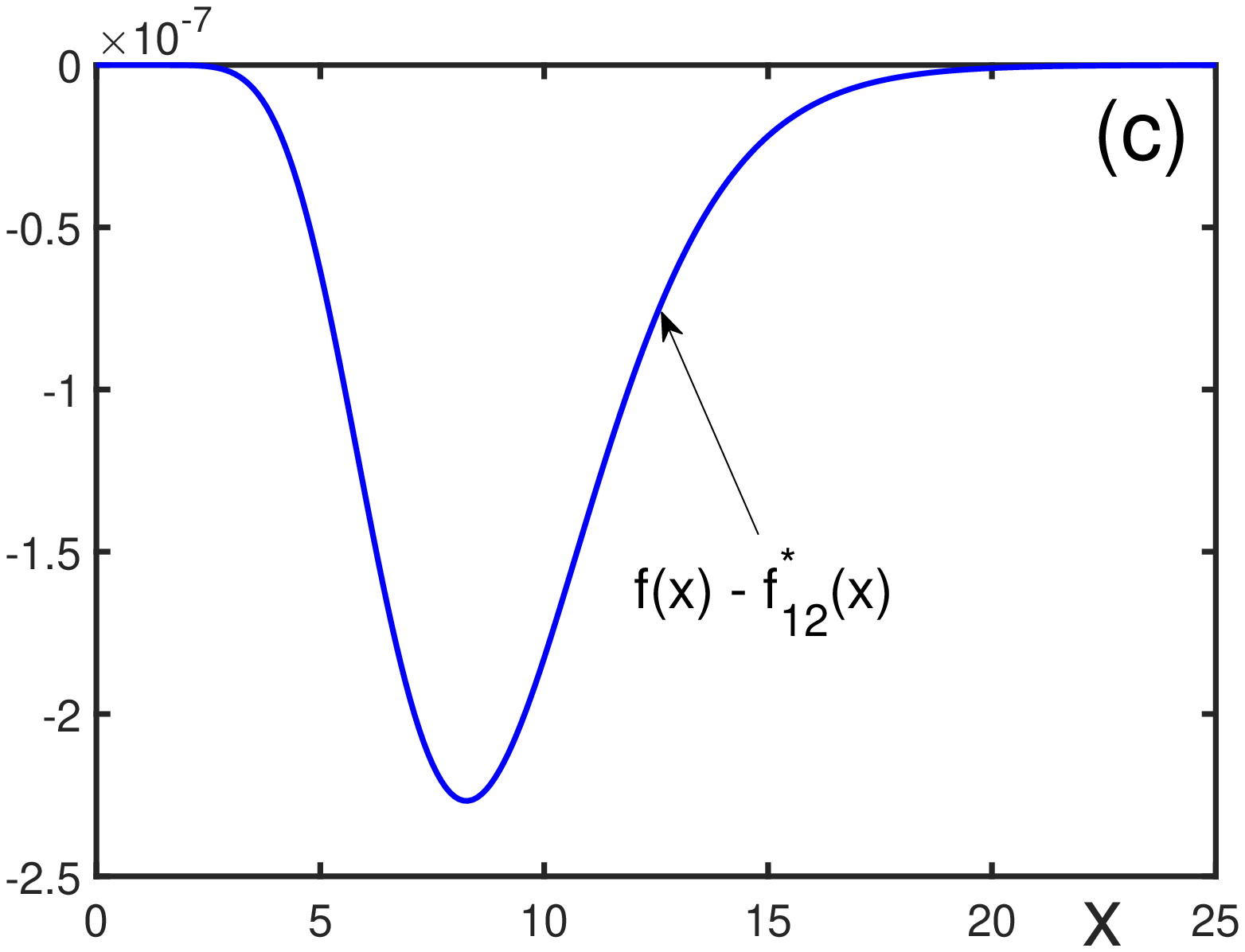} } }
\vskip 3mm
\caption{Difference between the Fermi-Dirac distribution (\ref{47}) and its
self-similar factor approximants: (a) $f_F(x)-f_4^*(x)$; (b) $f_F(x)-f_8^*(x)$;
(c) $f_F(x)-f_{12}^*(x)$. 
}
\label{fig:Fig.5}
\end{figure}

\newpage

\begin{figure}[ht]
\centerline{ \hbox{ 
\includegraphics[width=8cm]{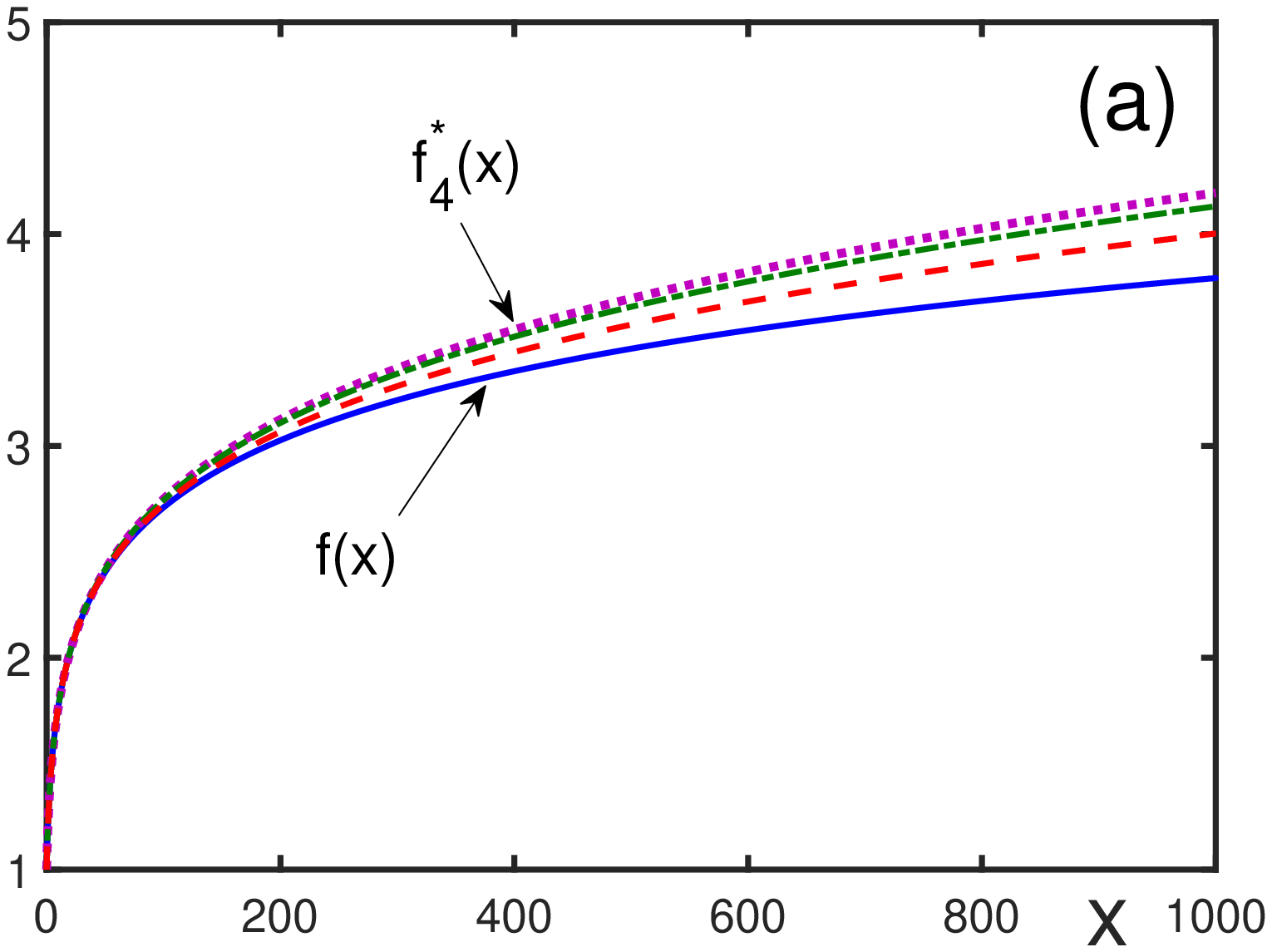} \hspace{1cm}
\includegraphics[width=8cm]{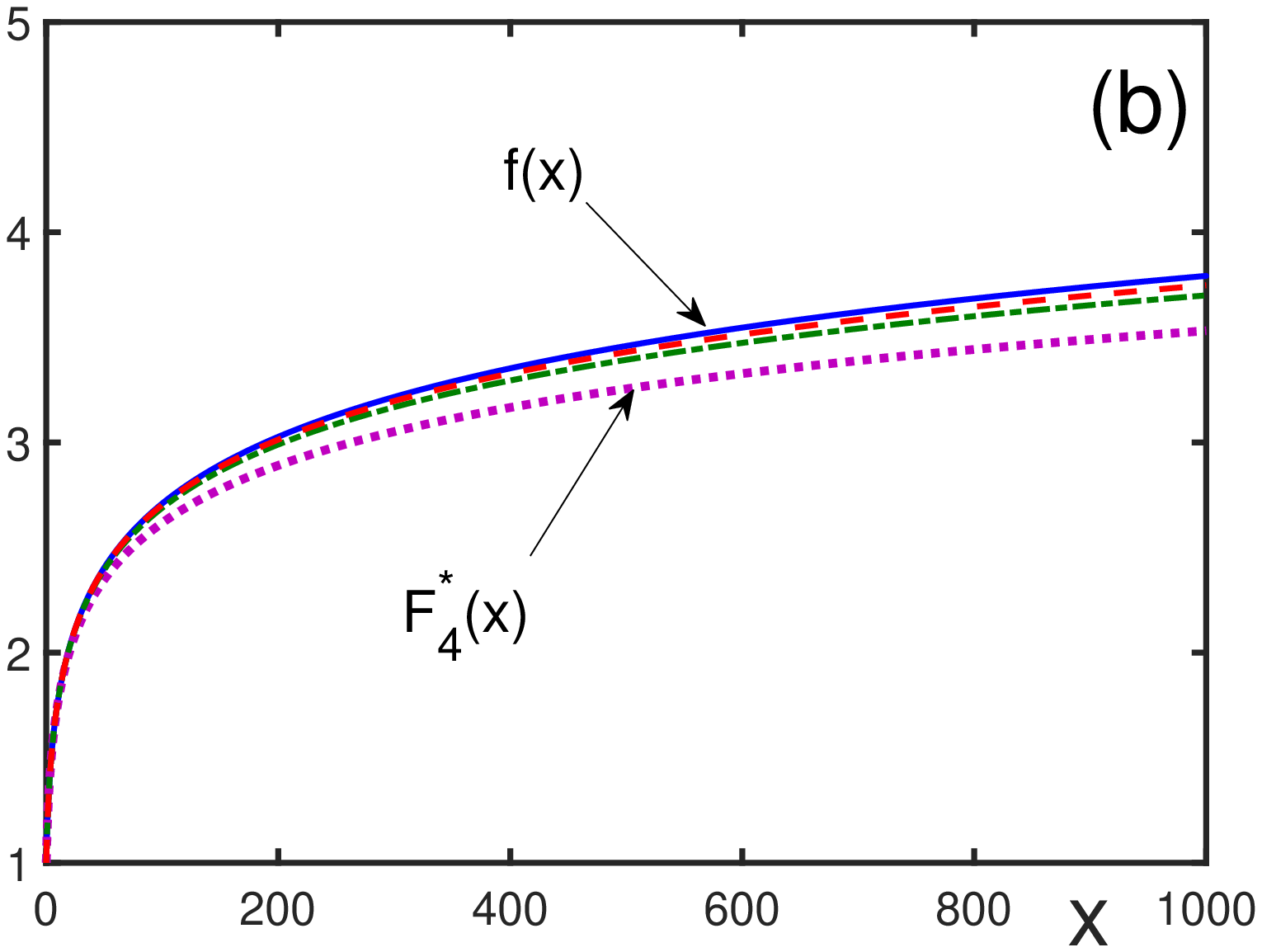}  } }
\vskip 3mm
\caption{(a) Function (\ref{49}) (solid line) and its self-similar factor 
approximants obtained from the direct series (\ref{51}): $f_4^*(x)$ (dotted line), 
$f_8^*(x)$ (dash-dotted line), and $f_{12}^*(x)$ (dashed line). (b) Function (\ref{49}) 
(solid line) and its self-similar factor approximants obtained from expansion (\ref{53}):
$F_4^*(x)$ (dotted line), $F_8^*(x)$ (dash-dotted line), and $F_{12}^*(x)$ (dashed line).
}
\label{fig:Fig.6}
\end{figure}

\end{document}